\documentclass[aps,tightenlines,twocolumn,prd,nofootinbib,
superscriptaddress,showpacs,preprintnumbers,eqsecnum,floatfix]{revtex4-1}
\def\bea{\begin{eqnarray}}
\def\be{\begin{equation}}
\def\ee{\end{equation}}
\def\eea{\end{eqnarray}}
\usepackage[usenames,dvipsnames]{color}

\usepackage{graphicx}
\usepackage{amsmath}
\usepackage{amssymb}
\usepackage{bbold}
\usepackage{bm}
\usepackage{slashed}
\usepackage{float}
\usepackage{psfrag}
\usepackage{footnote} 
\usepackage{xcolor}
\usepackage{textcomp}
\usepackage{cancel}
\usepackage{mathtools}
\begin{document}
  \preprint{JLAB-THY-21-3347}

\title{Removal of  singularities from the covariant spectator theory }

\author{Franz Gross }
 \affiliation{ Thomas Jefferson National Accelerator Facility (JLab), Newport News, Virginia 23606, USA}
 \affiliation{College of William and Mary, Williamsburg, Virginia 23188,
USA}

\email{gross@jlab.org}

\date{\today}
 \begin{abstract}
 A modification of the one boson exchange (OBE) kernel for the covariant spectator theory (CST) is presented and discussed.   When applied to the scattering of two identical particles, the previously used kernels either introduced spurious singularities, or removed them in an ad-hoc way.  The new modification not only removes these singularities, but also maintains the convergence  of the two-body  CST equation (sometimes called the Gross equation) when used to describe the scattering of two identical scalar particles. 

 \end{abstract}


\maketitle

\section{Introduction, History and background} \label{sec:Intro}

\subsection{Brief overview of this paper}

The covariant spectator theory (CST), formulated in Minkowski space, has enjoyed many successes during its long, over 50 year history.  One of its principal drawbacks is the presence of unphysical singularities that arise in the treatment of identical particles. The problem with these singularities is not a practical one, but rather a theoretical one: while most numerical calculations are insensitive to their presence, they are an unpleasant sign that something might be missing, and methods previously used for treating them have little justification, leaving room for doubt that the physics is fully under control. The main purpose of this paper is to discuss these singularities, and to propose and justify a new method for removing them.  

Mathematical details presented in this paper are limited to the simple scalar $\phi\psi^*\psi$ interaction (where the boson $\phi$ particles will have the mass $\mu$ of the pion in our numerical examples, but could be any spin zero particle, and the $\psi$ particles will be referred to as ``nucleons'' because they are a stand-in for nucleons or quarks, even though here they have spin zero).  This example is easy to examine in detail, and many of the conclusions obtained can be straightforwardly extended to more general theories with spin and isospin. 

In addition to this first section, this paper is divided into four other sections and a few short appendices with some details.  Sec.~\ref{sec:two}  looks at the lowest order one boson exchange (OBE) mechanism where the singularities first appear when the equations are used to describe the scattering of identical particles, and introduces the new method for removing them.  Sec.~\ref{sec:three} shows how these singularities affect the convergence of the CST by examining the next higher order kernels: the fourth order subtracted box and crossed box diagrams.  A strong motivation for use of the CST is the cancellation theorem, which states that  the higher order kernels describing the  scattering of nonidentical scalar particles cancel when one of the nucleon masses approaches infinity, leaving the OBE kernel to give the exact result in that limit.  This theorem is violated when the diagrams are symmetrized for the description of the scattering of identical particles, and  Sec.~\ref{sec:four} shows how the new method for removing singularities also improves the cancellation and almost restores the cancellation theorem, justifying its introduction.  Finally, conclusions are presented in the last Sec.~\ref{sec:conclusions}.    

Those familiar with the CST might prefer to skip parts of this introductory section, which includes subsections describing the assumptions that underly the use of the CST, a brief review of the early history of the CST, a very brief review of its major applications, and a general discussion of the different types of CST equations that describe (i) the scattering of nonidentical particles (one channel),  (ii) the scattering of identical particles with exchange symmetry (two channel) and (iii) the pion as a bound state of a $q\bar q$ pair, that requires four channels in order to maintain both charge conjugation invariance and spontaneously broken chiral symmetry.

\subsection{Underlying assumptions}

In hadronic physics, where a separation of scales is possible,   it is often assumed that the exchange of bosons is a long range force that can be correctly calculated by summing all ladders and crossed ladders (referred to as ``generalized'' ladders).  In this picture, vertex corrections and self energies are short range effects that may be treated through the introduction of phenomenological form factors, self energies, and dressed masses.   For these reasons, the summation of generalized ladders, and the Bethe-Salpeter (BS) and CST integral equations that are viewed as a way to sum these generalized ladder diagrams in closed form, are the focus of this paper.     

We also know that the existence of bound state poles requires an infinite sum of generalized ladders, even when the coupling is small.  And if the coupling is large and an infinite sum will not converge, integral equations provide solutions believed to be meaningful definition of this sum when the series they generate diverges.  

 \subsection{Early History}
 
The idea of putting a spectator on-shell (a covariant generalization of the nonrelativistic impulse approximation) was first introduced in a dispersion theory calculation of the deuteron form factors, published in 1964 \cite{Gross:1964mla,Gross:1964zz}.  This was before field theory had seen its revival and before QCD was widely known to the hadronic physics community.  

Briefly, the nonrelativistic impulse approximation for the deuteron form factors  treats one of the two nucleons in the deuteron as a spectator, with exchange current contributions treated separately and often regarded as a correction.   When this idea is applied to a relativistic calculation of the form factors using dispersion theory, it is found that the dominant contributions come from  the  region where the anomalous cut forces the spectator nucleon to be on-shell; a very nice covariant generalization of the spectator concept.  Extending this calculation requires a deuteron wave (or vertex) function which satisfies a relativistic bound state equation with one of the nucleons restricted to its mass shell.  This equation (sometimes known as the Gross equation) was introduced in 1969 \cite{Gross:1969rv}. 

The mid 1960's saw the development of many new ideas for how to solve the strong interactions nonpertubatively.  In 1966 Weinberg \cite{Weinberg:1966jm} suggested treating dynamics at  infinite momentum.  At about the same time Greenberg \cite{Greenberg:1965fw} introduced the N-quantum approximation, which he later realized \cite{Greenberg:2010wx} enjoys a close connection to the spectator equations.  A variety of quasi-potential equations were also introduced \cite{Woloshyn:1974wm}.  It wasn't until the 1980's that light front field theory \cite{Lepage:1980fj} developed  as a modern approach to hadronic physics. 

When applied to the generalized ladder (sum of all ladder and crossed ladder diagrams) approximation to the scattering of scalar, non-identical particles, with an interaction Lagrangian $\phi\chi^*\chi +\phi\psi^*\psi$ (where the $\chi$ particles will also  be referred to nucleons),  the 1969 paper showed that there existed a remarkable cancellation at fourth order.  In particular, when the mass of one of the particles approaches infinity, the irreducible fourth order kernel vanishes, so the exact result is the iteration of the OBE interaction.    [The irreducible part is the sum of the subtracted box diagram (the part of the box that remains when the iteration of the OBE is subtracted) and the crossed box.]   Later \cite{DRY,Gross:1993zj}
 it was shown that  the sum of {\it all\/} generalized ladder diagrams is well approximated using a CST scattering equation with {\it only a single\/} $\phi$ boson exchange kernel!  

While scalar theory cannot be used to describe the interactions of most physical systems, the cancellation theorem still leads to simplifications of the kernels needed to represent the generalized ladder sum for $NN$ scattering \cite{Gross:1982nz}.

\subsection{Applications of the CST}

The CST has been used to describe many systems.  This summary includes a very brief review of only five topics. Some other areas that have been studied are listed at the end of this subsection.
\vskip0.05in

\noindent{\bf Hydrogen-like atoms, muonium, and positronium:}  The earliest success of the CST was in the description of the hydrogen-like atoms, where the bound state CST equation reduces to a Dirac equation with an effective potential \cite{Gross:1969rv,Lepage:1977gd,Lepage:1978hz}.  These are ideal systems for applications of the CST \cite{Drell:1978ui}, which can be used to organize the perturbative corrections in bound state calculations \cite{Eides:1988ab}. Later, a nonrelativistic alternative was developed  that simplified the analytic calculations of the many integrals \cite{Caswell:1985ui}.  
\vskip0.05in

\noindent{\bf $NN$ scattering and deuteron structure:}  The  CST has been used to calculate the structure of the deuteron and predict $np$ scattering \cite{Gross:1972ye,Buck:1979ff,Gross:1991pm,Gross:2008ps,Gross:2010qm}.  The latest results \cite{Gross:2008ps,Gross:2010qm}, using the Model C described below, give an excellent fit to the $np$ phase shifts below lab energies of 350 MeV, with as few as 15 parameters.  Here the OBE model is justified by the cancellation theorem; for nucleons (with their spin an isospin) the cancellations are incomplete, but there are indications that, after the cancellations, what remains will contribute to the broad $\sigma$ resonance exchange \cite{Gross:1982nz,Pena:1996tf}.
\vskip0.05in

\noindent{\bf Deuteron form factors:} At first it was unclear how to calculate electromagnetic interactions of a CST bound state, but a paper with D.~O.~Riska \cite{Gross:1987bu} shows how to do this in  a gauge invariant manner.    Some early calculations did not have realistic relativistic wave functions, and used the general formalism to estimate relativistic corrections to the form factors \cite{Casper:1967zz,Arnold:1979cg}.    The first dynamically consistent calculation of the three form factors was done using Model B \cite{VanOrden:1995eg} (described below).  Later, using Model C and a momentum dependent $\sigma NN$ coupling,  the interaction current generated by this momentum dependence was determined  \cite{Gross:2014zra}, and precision calculations of the deuteron static moments and the form factors  we completed \cite{Gross:2014zra,Gross:2014wqa,Gross:2014tya,Gross:2019thk}.
None of the recent calculations use the isoscalar $\rho\pi\gamma$ (or $\sigma\omega\gamma$) exchange currents, which are probably small, very model dependent, and not required by current conservation  \cite{Ito:1993au}.

\vskip0.05in

\noindent{\bf Three-body bound states and form factors:}  CST equations have been derived for the treatment of three-body bound states \cite{Gross:1982ny,Stadler:1997iu} and the correct way to normalize a relativistic CST three-body bound state wave function determined \cite{Adam:1997rb}.  Using an early family of $np$ models, A.~Stadler solved the $3N$ CST equations for the triton and found that the momentum dependence of the $\sigma NN$ coupling that gives the best fit to the $np$ scattering phase shifts also predicts the correct triton binding energy \cite{Stadler:1996ut}!  Remarkably, this seems to be a robust feature of the CST, and was found to be  true also for the 2008  precision fit to the $np$ scattering data \cite{Gross:2008ps}.   The form of the CST three-body currents has been derived \cite{Kvinikhidze:1997wp,Gross:2003qi} and preliminary calculations of the three body form factors completed \cite{Pinto:2009dh,Pinto:2009jz}.
Experience with three-body equations suggests that it might be possible use the CST to convert nonrelativistic n-body equations of the AGS type \cite{AGS} into relativistic n-body equations, but this has not been done.

\vskip0.05in

\noindent{\bf Mesons as $q\bar q$ bound states:}   The CST is well suited to a study of $q\bar q$, $q\bar Q$, and $Q\bar q$ bound states (referred to collectively as $q\bar q$ states).  At first it would seem that placing a quark (or antiquark) on shell would be incompatible with the need for quark confinement, but three-dimensional nature of the CST makes it possible to construct a relativistic kernel that reduces to linear confinement (the flux tube)  in the nonrelativistic limit, and does not allow the quarks to escape the interaction region \cite{Gross:1991te}.   If $m_b$ is a $q\bar q$ bound state, it is even possible to show that the vertex function  $m_b\to q\bar q$ predicted by the CST equation with confinement is zero when both quarks are on-shell \cite{Savkli:1999me}.  Early results showed that it was possible to obtain a reasonable description of the light mesons \cite{Gross:1991pk,Gross:1994he}, and recently elegant calculations by the Portuguese group successfully fit the spectrum of heavy and heavy-light mesons \cite{Leitao:2014jha,Leitao:2016bqq,Leitao:2017mlx}.  One motivation for this paper is to address one issue that must be solved before the Portuguese calculations can be extended to the light mesons.
\vskip0.05in

\noindent{\bf Quark-self energies:}  Most recently, partly in preparation for a calculation of the light meson spectrum, the Portuguese group and I have turned to studies of the quark self-energy   \cite{Biernat:2012ig,Biernat:2013fka,Biernat:2013aka,Biernat:2014xaa,Biernat:2015xya,Biernat:2018khd}.  The CST  is not naturally designed for self-energy calculations, and they have turned out to be a challenge. This effort has lead to new insights, and is one of the motivations for this paper.  

\vskip0.05in

\noindent{\bf Other systems} that have been studied using the CST include pion-nucleon scattering \cite{Gross:1992tj,Surya:1995ur}, relativistic proton nucleus scattering \cite{Gross:1988kq,Gross:1989kx,MaungMaung:1991qv,Wallace:1994ux,Wallace:1995ze}, relativistic nuclear matter \cite{Anastasio:1980dz}, the EMC effect \cite{Gross:1991pi,Liuti:1995uf}, and non-perturbative studies using the Feynman-Schwinger technique \cite{Savkli:1999rw,Savkli:1999ui,Gross:2001ha,Savkli:2002fj,Savkli:2004bi}.  

\begin{figure*}
 \centering
 \includegraphics[width=6in]{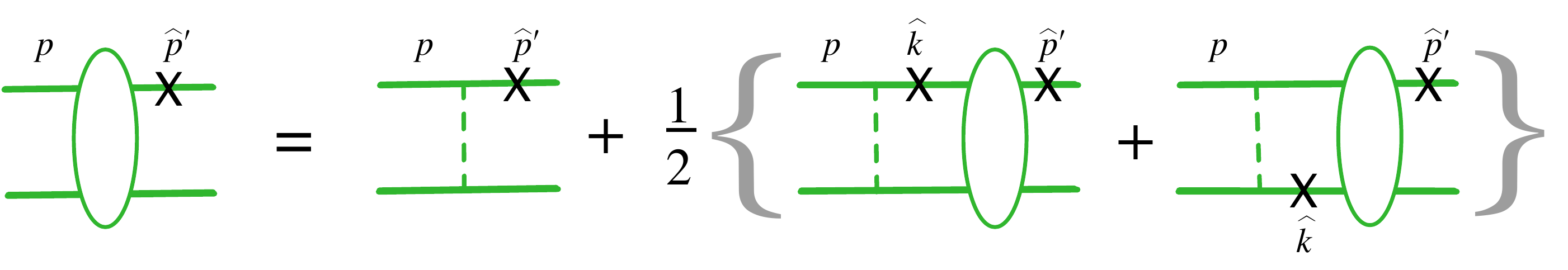}
 \caption{Graph showing the off-shell two-channel CST equation (\ref{eq:2channeloffshell}).  The momenta of the unlabeled particles are always $P$ minus the momenta of the labeled particles.  Note that the momenta of the internal particles in the last diagram differ from the usual definition.}
 \label{fig:2channel}
\end{figure*}

\begin{figure*}
 \centering
 \includegraphics[width=6in]{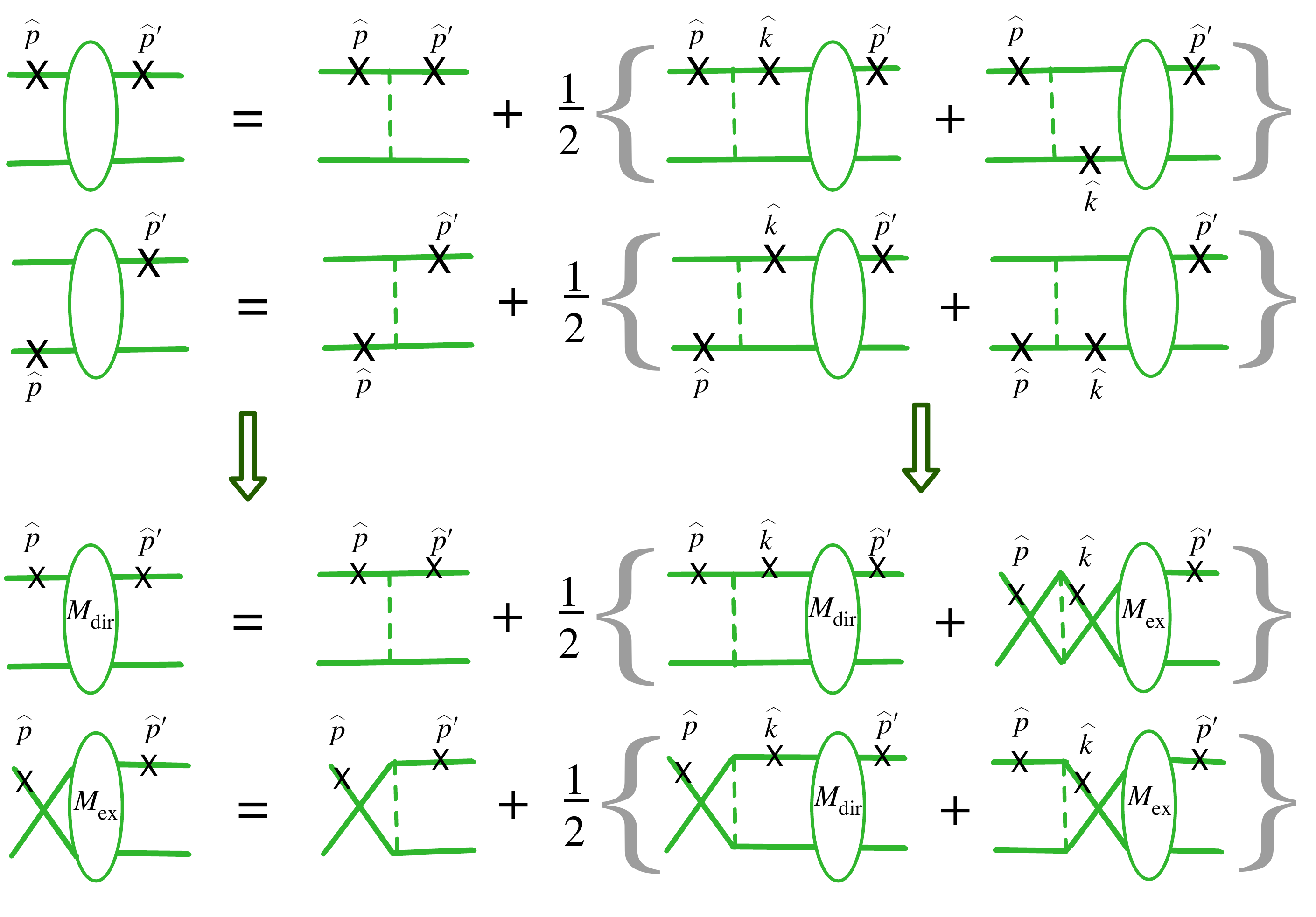}
 \caption{Top two rows are the two-channel coupled equations constructed from Fig.~\ref{fig:2channel} by putting different final state particles on-shell  (accompanied by a corresponding change in the labeling of the momenta).  The lower two rows show how nucleon lines in the upper diagrams can be exchanged (always maintaining the same internal connections so that numerical evaluations are unaffected), leading to  Eq.~(\ref{eq:2channel}).  }
 \label{fig:2channel2}
\end{figure*}

\subsection{One-channel CST equation}

In this paper the four-momenta of two interacting particles 1 and 2 (referred to as nucleons even though they could be quarks, and have spin zero) will be denoted
\bea
p_1&=&p
\nonumber\\
p_2&=&P-p\, ,
\eea
where $P$ is the total four-momentum of the pair, conserved in the interactions, and $p$ (or $k$, etc.)~is the momentum of particle 1 (beware that in many previous references, $p$ or $k$ was used to denote the {\it relative\/} momentum of the pair).  When the particles are nonidentical, the CST prescription places the heavier particle, which I choose to be particle 1 with mass $M$, on-shell.  When the particle is on-shell, $p$ is replaced by $\hat p$, where
\bea
\hat p = \{E_p, {\bf p}\}=\{\sqrt{M^2+p^2},{\bf p}\}\,.
\eea
(Note that $p$ will sometimes represent the magnitude of ${\bf p}$ and sometimes the four-vector; the distinction should be clear from the context.)

The CST scattering equation, for two nonidentical scalar particles with particle 1 on shell, has the form
\bea
{\cal M}(\hat p,\hat p';P)&=&{V}(\hat p, \hat p';P)
\nonumber\\
&&+\int_k {V}(\hat p,\hat k;P)G(\hat k,P){\cal M}(\hat k, \hat p';P),\qquad
\label{eq:scattering}
\eea
where ${\cal M}$ is the two-body scattering amplitude  resulting from the solution of the equation, ${G}(\hat k,P)$ is the two-nucleon propagator, which in the rest system $P=\{W,{\bf 0}\}$  
is 
\bea
G(\hat k,P) = \frac1{W(2E_k-W-i\epsilon)} , \label{eq:G0}
\eea
and the integral over the three-momentum $k$ is 
\bea
\int_k=\int\frac{d^3k}{(2\pi)^3 2E_k}  .\label{eq:intk}
\eea

This equation can be used to {\it define\/} the amplitude when {\it  both\/} particles are off shell in the final state:
\bea
{\cal M}(p, \hat p';P)&=&{V}(p, \hat p';P)
\nonumber\\
&&+\int_k{V}(p,\hat k;P)G(\hat k,P){\cal M}(\hat k, \hat p';P).\qquad
\label{eq:scattering2}
\eea
Once ${\cal M}(\hat p, \hat p';P)$ has been found by solving (\ref{eq:scattering}), the off-shell amplitude ${\cal M}(p, \hat p';P)$ can be computed from  (\ref{eq:scattering2}), provided the kernel ${V}$ is known off shell.


\subsection{Two-channel CST equations}

The original Ref.~\cite{Gross:1969rv} discussed a one-channel equation.   Later, when  the CST was applied to $NN$ scattering \cite{Gross:1991pm}, where the two particles are identical and the scattering amplitude must be symmetric or antisymmetric under particle interchange (the generalized Pauli principal), it was necessary to introduce a two-channel equation to insure that this symmetry emerges automatically from the solutions.

The construction of a two channel equation that will give the symmetry automatically was derived and discussed in detail in Ref.~\cite{Gross:1991pm}.  This method introduces the singularity that will be discussed in the next section.  [An alternative method with no singularity was discussed in detail in Appendix B of Ref.~\cite{Gross:2008ps}, but does not automatically give symmetric solutions.]

The derivation of the correct equation is shown diagrammatically Figs.~\ref{fig:2channel} and \ref{fig:2channel2}.  Since the two particles are identical, a symmetric result is obtained by averaging over the contributions from particle 1 on-shell  plus  particle 2 on-shell (for convenience the momentum of both internal on-shell particles are labeled by the four vector $\hat k$).   If particle 1 is on-shell in the initial state, and {\it both\/} particles are off-shell in the final state, Fig.~\ref{fig:2channel} illustrates the equation, which for a spin zero system is written
\begin{widetext}
\bea
{\cal M}(p, \hat p';P)= V(p, \hat p';P) +\frac12\int_k \Big\{V(p, \hat k; P)G(\hat k,P) {\cal M}(\hat k, \hat p'; P)+ V(p, P-\hat k; P)G(\hat k,P) {\cal M}(P-\hat k, \hat p'; P)\Big\} .\qquad \label{eq:2channeloffshell}
\eea
This can be written as two coupled equations for the amplitudes.  Using the notation 
\bea
{M}_{\rm dir}(\hat p, \hat p';P)&=&{\cal M}(\hat p, \hat p';P)
\nonumber\\
{M}_{\rm ex}(\hat p, \hat p';P)&=&{\cal M}(P-\hat p, \hat p';P)
\eea
and assuming that the kernel has the following properties
\bea
{V}_{\rm dir}(\hat p, \hat p';P)&\equiv&{V}(\hat p, \hat p';P)={V}(P-\hat p, P-\hat p';P)
\nonumber\\
{V}_{\rm ex}(\hat p, \hat p';P)&\equiv&{V}(P-\hat p, \hat p';P)={V}(\hat p, P-\hat p';P) \qquad \label{eq:defV}
\eea
these equations are
\bea
{M}_{\rm dir}(\hat p, \hat p';P)&=&{V}_{\rm dir}(\hat p, \hat p';P)
+\frac12\int_k \Big\{{V}_{\rm dir}(\hat p,\hat k;P)G(\hat k,P){M}_{\rm dir}(\hat k, \hat p';P)+{V}_{\rm ex}(\hat p,\hat k;P)G(\hat k,P){M}_{\rm ex}(\hat k, \hat p';P)\Big\}
\nonumber\\
{M}_{\rm ex}(\hat p, \hat p';P)&=&{V}_{\rm ex}(\hat p, \hat p';P)
+\frac12\int_k \Big\{{V}_{\rm ex}(\hat p,\hat k;P)G(\hat k,P){M}_{\rm dir}(\hat k, \hat p';P)+{V}_{\rm dir}(\hat p,\hat k;P)G(\hat k,P){M}_{\rm ex}(\hat k, \hat p';P)\Big\} ,\qquad
\label{eq:2channel}
\eea
\end{widetext}
and the notation and labeling of momentum is shown in Fig.~\ref{fig:2channel2}.  Note that the first equality in each line of (\ref{eq:defV}) can be regarded as a definition, but the second equality is an assumed property of the kernel.  The OBE kernels I will use are a function of $q^2$, and satisfy this property.

Multiplying the second equation by $\eta=\pm1$,  averaging the two equations, and using the notation
\bea
\bar{V}(\hat p,\hat k;P)=\frac12\Big[V_{\rm dir}(\hat p,\hat k;P)+\eta V_{\rm ex}(\hat p,\hat k;P)\Big] . \qquad \label{eq:Vtotal}
\eea
for both $V$ and $M$, gives
\bea
\bar{M}(\hat p, \hat p';P)&=&\bar{V}(\hat p, \hat p';P)
\nonumber\\&&
+\int_k \bar{V}(\hat p,\hat k;P)G(\hat k,P)\bar{M}(\hat k, \hat p';P) .\qquad\quad
\eea
The solution of this equation automatically satisfies the symmetry condition 
\bea
\bar{M}(\hat p, \hat p';P)=\eta\, \bar{M}(P-\hat p, \hat p';P)
\eea
under particle interchange
\bea
\hat p \leftrightarrow P-\hat p\, .
\eea
Here  $\eta$ is a symmetry factor that will depend on particle spin or isospin.  This is the generalized Pauli principle, required of all realistic descriptions of identical particles, but,  because only one particle is on shell, it can only be satisfied in a CST equation by explicit construction.  In  the examples discussed in this paper, where the particles have spin and isospin zero, $\eta=1$, but I will keep the factor so that the separate contributions from the direct and exchanged diagrams can be tracked.

\subsection{Four-channel CST equations}

So far the four-channel CST equations have been used only to describe the $q\bar q$ structure of the pion.  In this case it is important to preserve charge conjugation symmetry and treat chiral symmetry breaking correctly, which requires maintaining the correct limit when the bound state pion mass, $m_\pi$, goes to zero.  Charge conjugation symmetry is handled much the same way as particle interchange symmetry.  For particle interchange, both particles are outgoing and positive energy mass-shell contributions from each are averaged.  For the $q\bar q$  state, one particle is an outgoing antiquark (or an incoming quark) which is placed on its negative (positive) energy mass shell, different in detail but not in principle.  This requires two channels.  However, in order to get the correct chiral limit when $m_\pi\to 0$, two more channels are needed because in this limit the positive and negative energy quark poles coalesce, and both must be considered.  This is discussed in detail in Refs.~\cite{Leitao:2017mlx} and \cite{Biernat:2013fka}.

With this background discussion completed, I now discuss the issue of singularities and their removal.

\section{OBE: Second order diagrams} \label{sec:two}

The shortcoming of previous calculations was the appearance of spurious singularities in the OBE exchange kernel $V_{\rm ex}$, or the way in which they were removed.  These singularities will be discussed in this section. 

 \subsection{Definition of the OBE kernel}
 
 In both the CST and the Bethe-Salpeter  (BS) theory, the OBE kernel of spin zero particles  is
 \bea
 V(q) = -\frac{g^2M\mu\,F_b(q^2)}{\mu^2-q^2-i\epsilon}\,, \label{eq:OBE}
 \eea
 where, as in Ref.~\cite{Gross:1969rv}, I have scaled the coupling constant  
 so that $g^2$ is dimensionless, and 
$F_b(q^2)$ is a boson form factor.  In an unregularized theory,  $F_b=1$. I use $V(q)$ to make the notation compact; hence 
\bea
V(q)\equiv V(-q).\label{eq:VqtoVmq}
\eea
In the remainder of this paper, this definition replaces the more general one used in Sec.~\ref{sec:Intro}   
 
 \begin{figure}[b]
\begin{center}
 \includegraphics[width=2.5in]{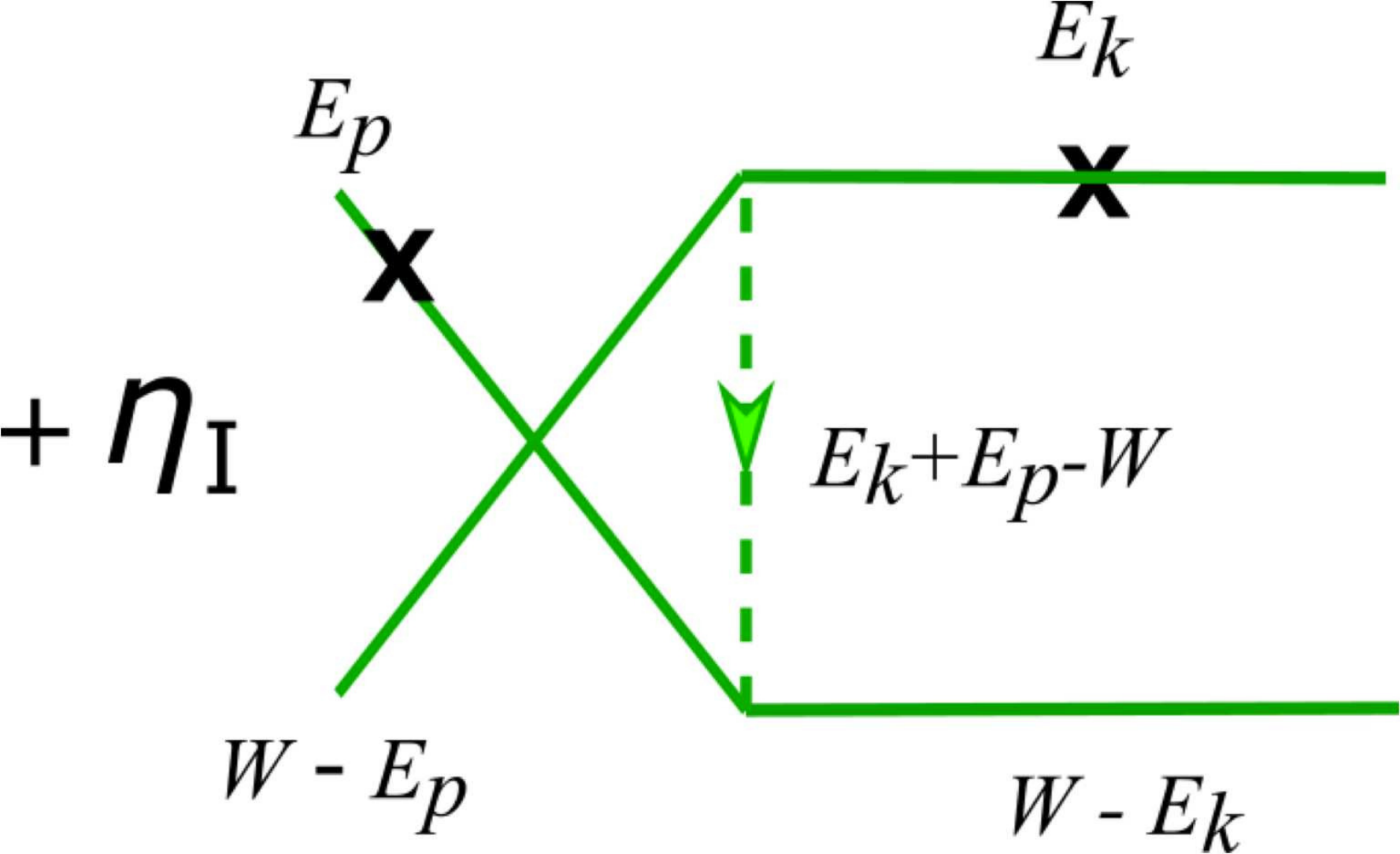}
\caption{OBE exchange diagram with particle 1 on shell in both the initial state and final states, and in this paper $\eta_I=\eta=1$.  }
\label{fig:1}
\end{center}
\end{figure}

 It is important to realize that the functional form of the OBE kernel is the same for both BS and CST.  The difference (which is significant) is the range of possible values that the four-vectors $p$ and $k$ can assume.  In the BS formalism,  these four vectors can assume {\it any\/} real values of all four components, while in the coupled CST equations (\ref{eq:2channel})  they may only be (in the cm frame)
 \bea
p&\to&\{E_p,{\bf p}\}\;{\rm or}\; \{W-E_p, -{\bf p}\}
 \nonumber\\
k&\to&\{E_k,{\bf k}\}\;{\rm or}\; \{W-E_k, -{\bf k}\}\, ,
 \eea 
 corresponding to the choices of  particle 1 on-shell in either the final or initial state, together with one of the interchanges (\ref{eq:defV}).   Only one interchange is needed, and my convention in this paper is to confine the interchange to the final state.  Since the CST limits  the domain of possible values of $p$ and $k$, symmetrization requires averaging the two possible on-shell values of $p$ (or $k$), whereas the BS does not require explicit symmetrization because all values of the momenta are among the  acceptable possibilities. 
 
 \subsection{Singularities in the exchange term}
 
The boson propagator of the exchange term with $q^2=(P-\hat p-\hat k)^2$ has poles at 
\bea
E_p+E_k\pm\omega_{-p}=W\equiv W_\pm\, , \label{eq:Wpm}
\eea
where this is written in terms of the boson on-shell energy\bea
\omega_{\pm p}^2=\mu^2+({\bf k}\mp{\bf p})^2\, .
\eea
The pole at $W_+\geq 2M+\mu$ is the  singularity that arises at energies when physical bosons can be produced, and is expected.   However, the singularity at $W_-$ is spurious because it has no physical origin.  It arises from a singularity that occurs when the off-shell baryon is ``unstable'';  such singularities should not appear in physical amplitudes.  In the remainder of this paper I will refer to these as ``instability'' (rather than ``spurious'') singularities.

\begin{figure}
 \centering
 \includegraphics[width=3in]{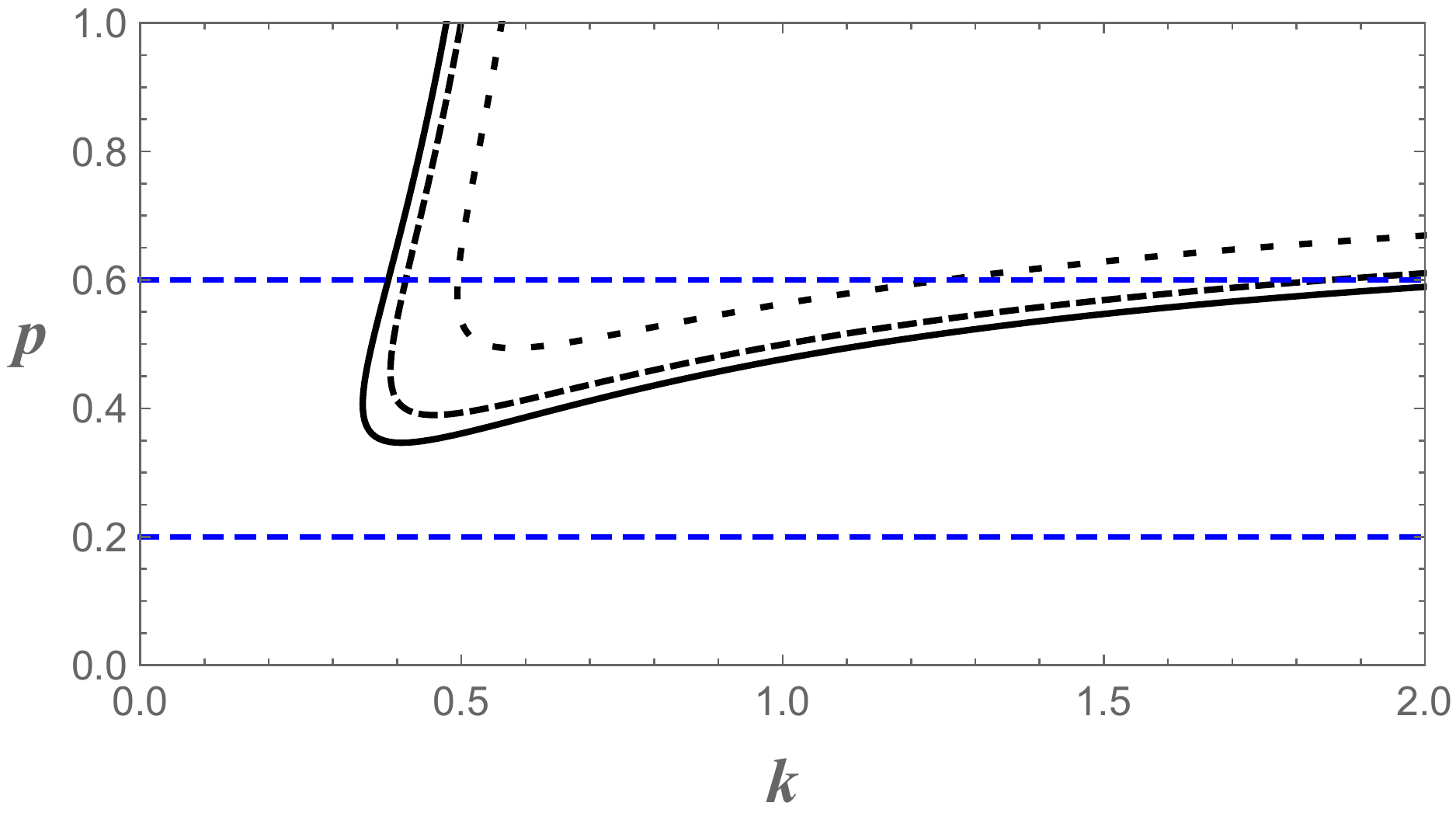}
 \caption{Contour lines showing the location of the singularity at $W_-$ in the $p$ and $k$ plane (for a fixed $z=-1$)  at three energies $W$: 1.978 (black), 2.02 (long-dashed black), and 2.1 (short-dashed black).  The two dotted blue lines are fixed values of $p=$ 0.2 and 0.6. Units are in the baryon mass. }
 \label{fig:contour}
\end{figure}

To explain this in more detail, focus on the lower vertex of Fig.~\ref{fig:1}.  
The on-shell physical baryon in the final state will have an energy larger than a physical boson with energy $\omega_{-p}$ and an {\it unphysical\/} baryon with  energy, $W-E_k$ when 
\bea
E_p \geq W-E_k +\omega_{-p}\, . \label{eq:instability}
\eea
The threshold for this instability is therefore at $W=W_-$.  For every value of $W$ and $z$, these ``instability'' singularities lie on a curve in the $p, k$ plane, symmetric about the diagonal line $p=k$.  Examples of these curves for three values of $W$ (with $z=-1$) are shown in Fig.~\ref{fig:contour}.   In all numerical examples, $\mu/M  = 0.139/0.939=0.148$, derived from the physical pion and nucleon masses.

In both the BS and CST approaches,  this instability does not appear when the initial state is on-shell (so that $W=2E_k$), because the inequality (\ref{eq:instability}) can never be satisfied.  It show this, first observe that, if $W=2E_k$, the inequality becomes
\bea
E_p-E_k\geq \omega_{-p}\, , \label{eq:onshellcase}
\eea
requiring that $p>k$.  Squaring both sides and rearranging some terms gives the condition 
\bea
2M^2-2E_pE_k+2pkz&\geq& \mu^2\, , \label{eq:ineq1}
\eea
where $z$ is the cosine of the angle between ${\bf p}$ and ${\bf k}$.
But this inequality cannot be satisfied because 
\bea
&&\frac{d}{dp}\big[2M^2-2E_pE_k+2pkz\big] = 2\Big[kz-\frac{p E_k}{E_p}\Big]
\nonumber\\
&&\qquad \leq  2p\Big[1-\frac{E_k}{E_p}\Big]\leq 0\, .
\eea
Hence, since the l.h.s. of  (\ref{eq:ineq1}) decreases with $p$, its largest value is at $z=1$ and $p=k$, where it is zero.  This completes the proof. 



Thus the instability first appears when the kernel is iterated, or, as required for applications to the three body problem, when the initial state must also be off-shell.  In my first work with J.~W.~Van Orden \cite{Gross:1991pm}, our Models B  kept these instability singularities and evaluated them using the principle value prescription.  Later, in work with A.~Stadler \cite{Gross:2008ps}, we introduced Models C which removed the singularities  by replacing the four-momentum of the exchanged meson with its absolute value
\bea
(p-\hat k)^2\to|(p-\hat k)^2|\, . \label{eq:precC}
\eea

Neither of these models is fully satisfactory, even though they both give finite results.  It is difficult to get numerically stable results with Model B, but more importantly, as will be seen below,  this model seriously  violates the cancellation  theorem in the region of the singularity.  It undermines the justification for using the CST.   Model C is not fully satisfactory because the absolute value introduces undesirable discontinuities in the derivatives of the kernel.  

\subsection{Cancelling the instability}

To solve these problems, I present a new CST model which uses a subtraction to cancel the instability.  
The motivation for the subtraction comes from study of the fourth order diagrams discussed below, and its advantage  is that  it smoothly preserve the analytic behavior of the kernels near the instability.  

\begin{figure}
 \centering
 \includegraphics[width=3in]{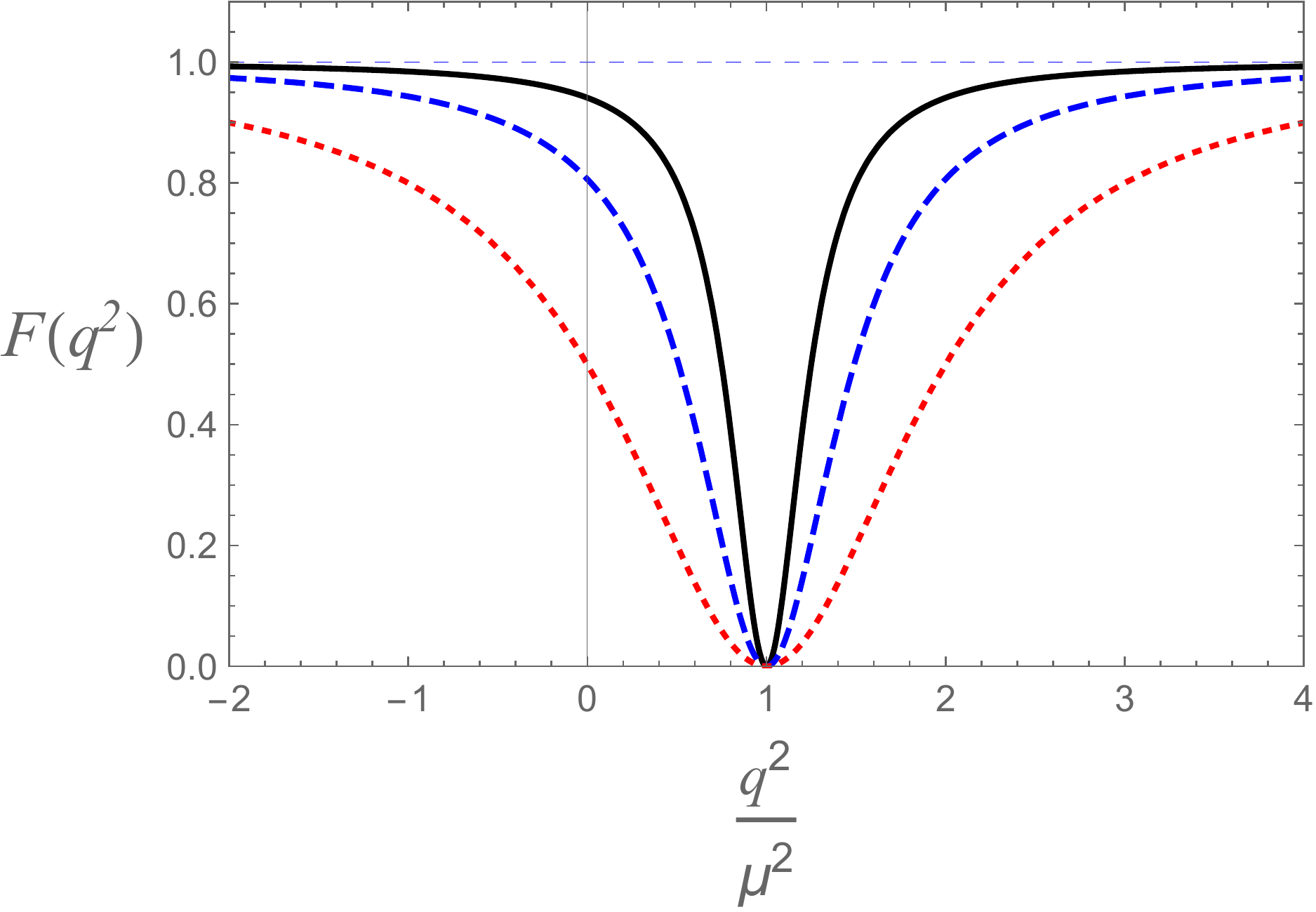}
 \caption{Form factor (\ref{eq:Fq2}) for $\lambda = 0.5$ (black), 0.7 (blue dashed), and 1 (red dotted). }
 \label{fig:Fq2}
\end{figure}

It is desirable that the subtraction (i) depend only on $q^2$ 
so that it will not introduce new electromagnetic exchange currents, (ii) vanish rapidly when $q^2\ne \mu^2$ so that the original CST kernel is preserved as much as possible, (iii) have no singularities on the real $q^2$ axis, and (iv)  be simple.  This leaves little freedom for the construction of the term, and an almost unique choice is 
\bea
V^{\rm D}(\hat k-\hat p)&=&-\frac{g^2M\mu}{\mu^2-q^2-i\epsilon}\Bigg\{1-\frac{\lambda_\mu^4}{\lambda_\mu^4+(\mu^2-q^2-i\epsilon)^2}\Bigg\}
\nonumber\\
&=&-\frac{g^2M\mu\,F_b(q^2)}{\mu^2-q^2-i\epsilon}\, , \label{eq:OBED}
\eea
where $\lambda_\mu=\lambda\mu$, with $\lambda$ is a dimensionless adjustable parameter, and the form factor is 
\bea
F_b(q^2)=\frac{\left[1-\frac{q^2}{\mu^2}-i\epsilon\right]^2}{\lambda^4+\left[1-\frac{q^2}{\mu^2}-i\epsilon\right]^2}\, .
\label{eq:Fq2}
\eea
Note the placement of the $i\epsilon$ associated with the boson mass; this seems unimportant now but will play an important role later.
The limit $\mu^2 \to 0$ (for {\it fixed\/} $\lambda$), gives $F_b(q^2) = 1$. This is a nice feature since the one-photon-exchange mechanism, which requires a $1/q^2$ singularity near $q^2=0$ in order to reproduce the correct Coulomb limit, is not modified.  For non-zero boson masses, $F_b(\mu^2)=0$, cancelling both the instability and production singularities.  Fig.~\ref{fig:Fq2} shows the  behavior of this form factor for three choices of $\lambda$.

\begin{figure*}[b]
 \leftline{\includegraphics[height=2.2in]{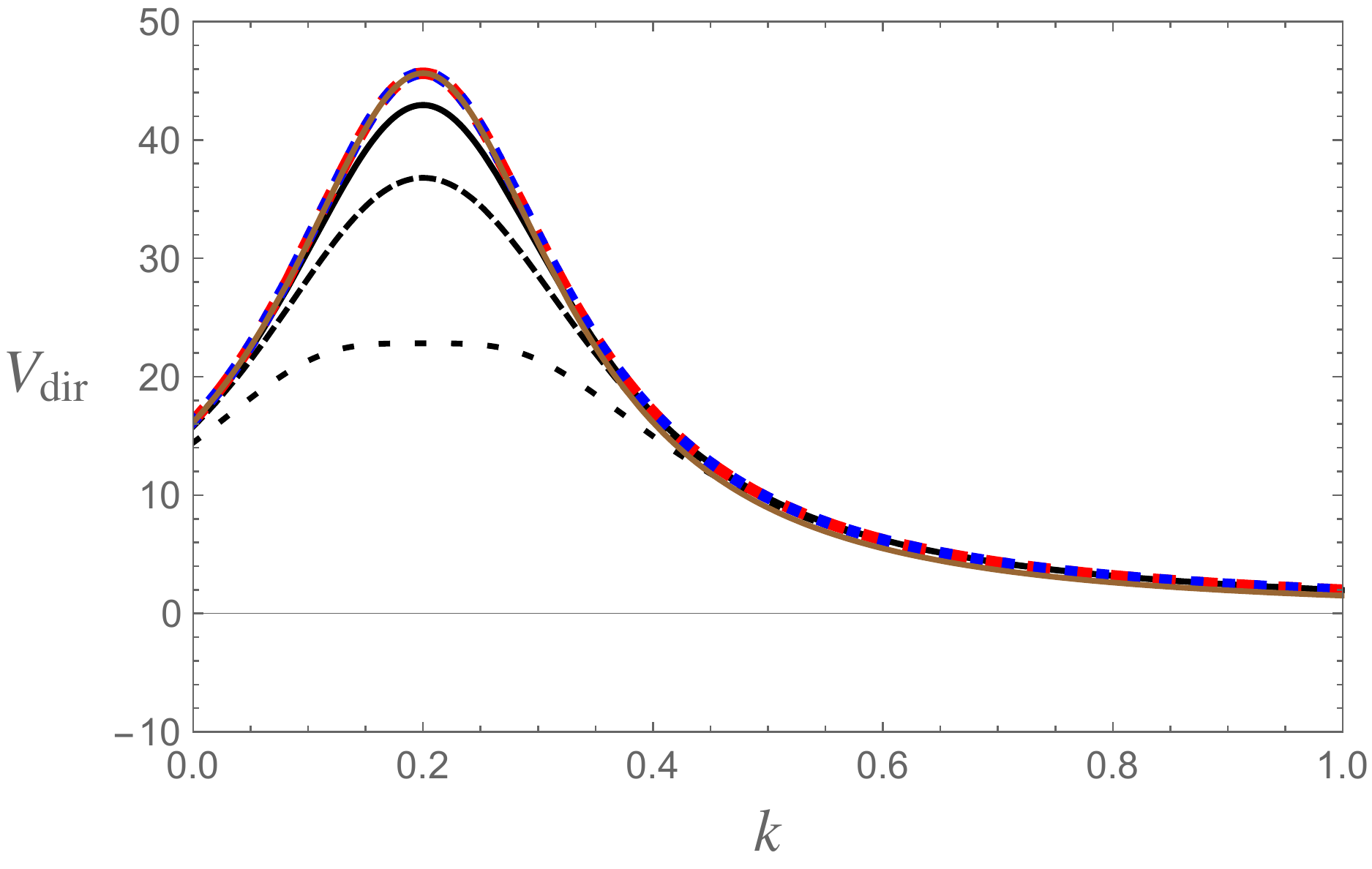}}
  \vspace{-2.2in}
\rightline{\includegraphics[height=2.2in]{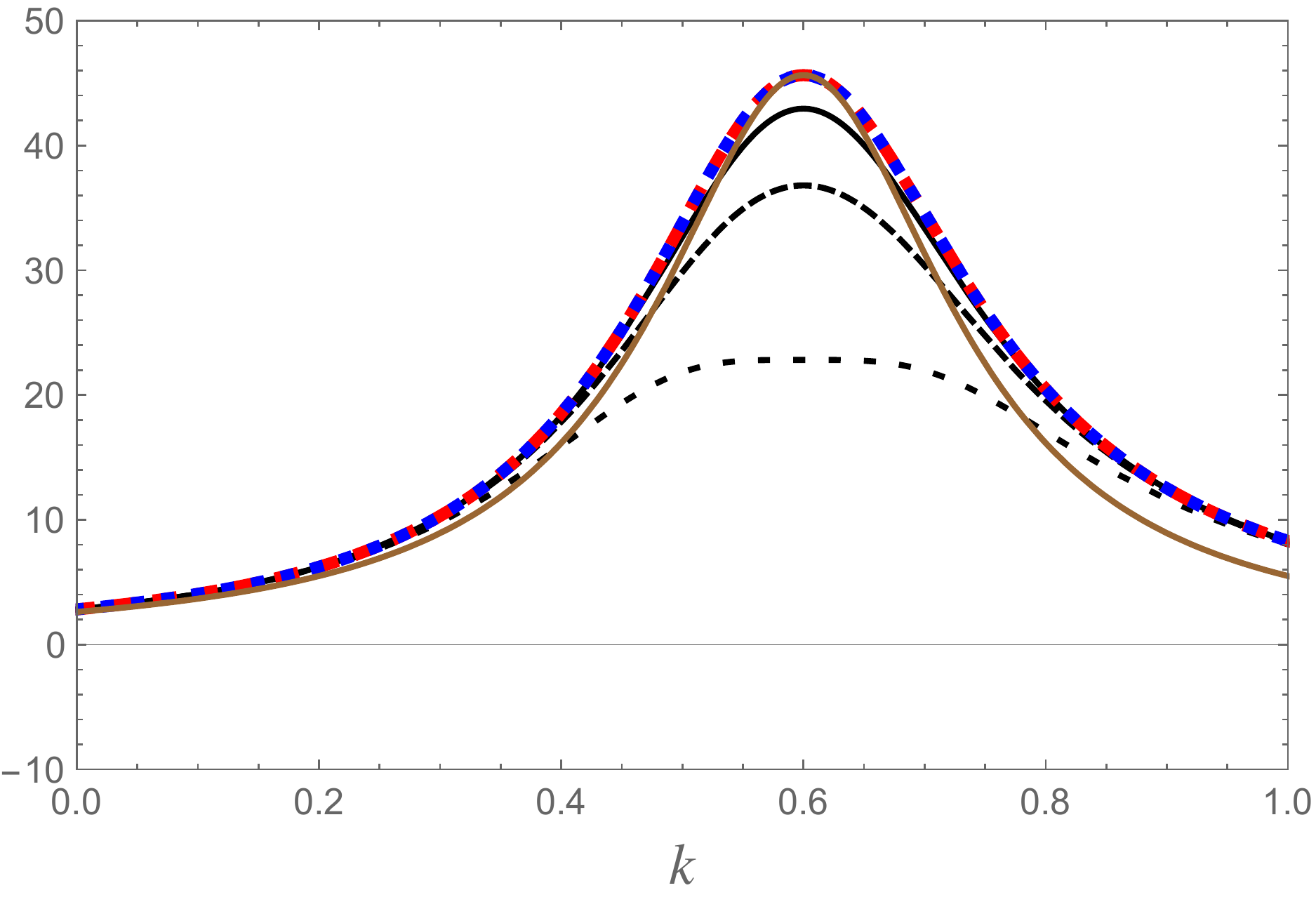}}
 \caption{{\footnotesize Left panel: lines for $z=1$, $p=0.2$ showing the direct kernel $V_{\rm dir}^{\rm D}$ as a function of $k$ for $\lambda=0.5$ (black), $\lambda=0.7$ (long-dashed black) and $\lambda=1$ (short-dashed black).  The other cases, also for $z=1$ and $p=0.2$,  are the nonrelativistic limit (brown), $V_{\rm dir}^{\rm B}$ (red dashed),  and $V_{\rm dir}^{\rm C}$ (blue dashed).  The dimensionless units are defined in the text.  Right panel: the same cases but for $p=0.6$.}}
 \label{fig:dir}
\end{figure*}

\begin{figure*}
 \leftline{\includegraphics[height=2.1in]{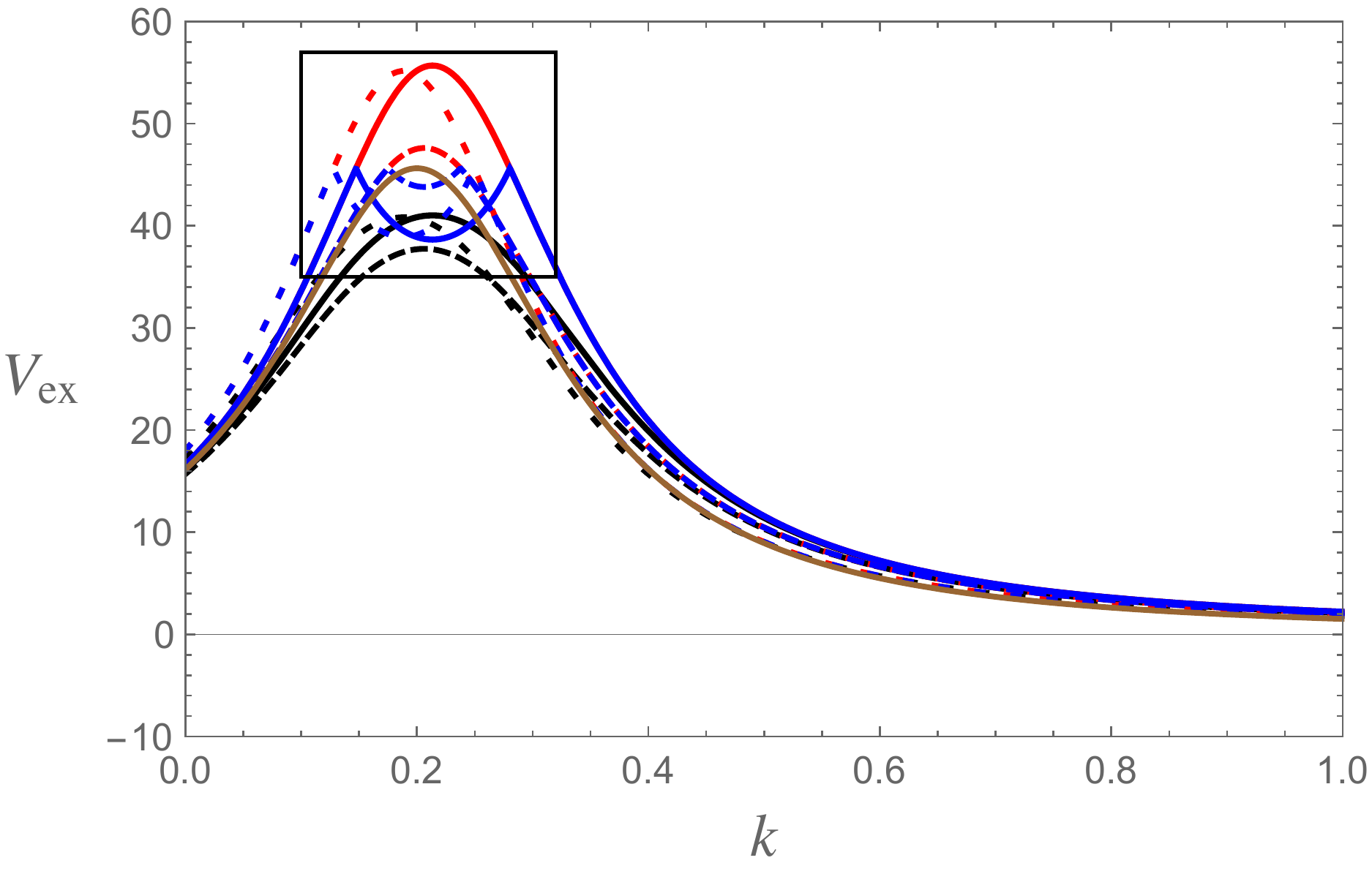}}
  \vspace{-2.1in}
\rightline{\includegraphics[height=2.1in]{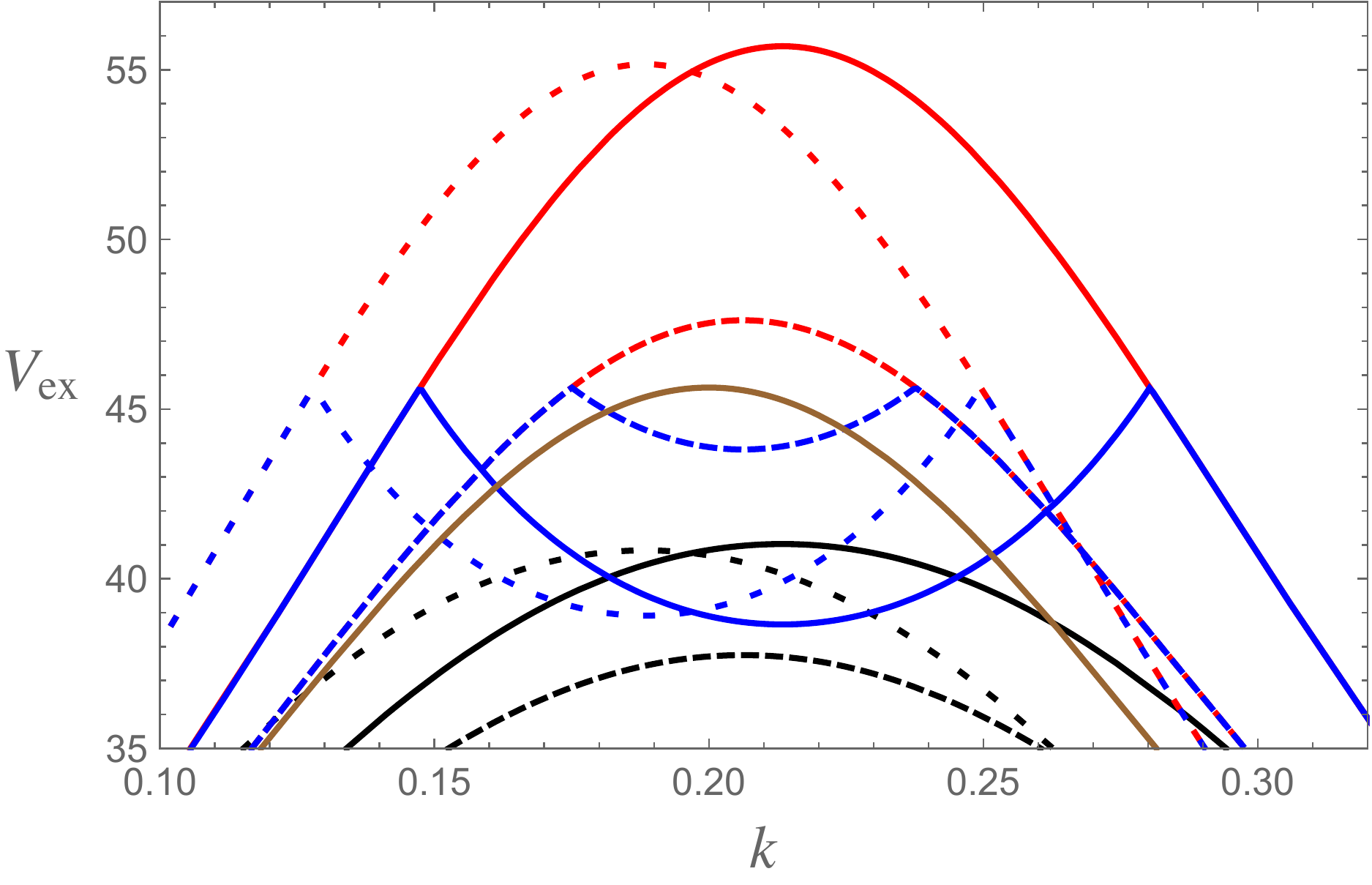}}
 \caption{{\footnotesize Left panel: lines for $z=-1$, $p=0.2$ showing the exchange kernel $V_{\rm ex}^{\rm D}$ with $\lambda=0.7$, as a function of $k$ for $W=1.978$ (black), $W=2.01$ (long-dashed black) and $W=2.1$ (short-dashed black), the singular kernel $V_{\rm ex}^{\rm B}$ as a function of $k$ for $W=1.978$ (red), $W=2.01$ (long-dashed red) and $W=2.1$ (short-dashed red),  and the regularized kernel $V_{\rm ex}^{\rm C}$ as a function of $k$ for $W=1.978$ (blue), $W=2.01$ (long-dashed blue) and $W=2.1$ (short-dashed blue). The nonrelativistic limit, independent of energy,  is shown for comparison (brown). The dimensionless units are defined in the text.  Right panel: the same cases on the expanded scale shown in the box in the left panel.}}
 \label{fig:exa}
\end{figure*}

\begin{figure*}
 \leftline{\includegraphics[height=2.2in]{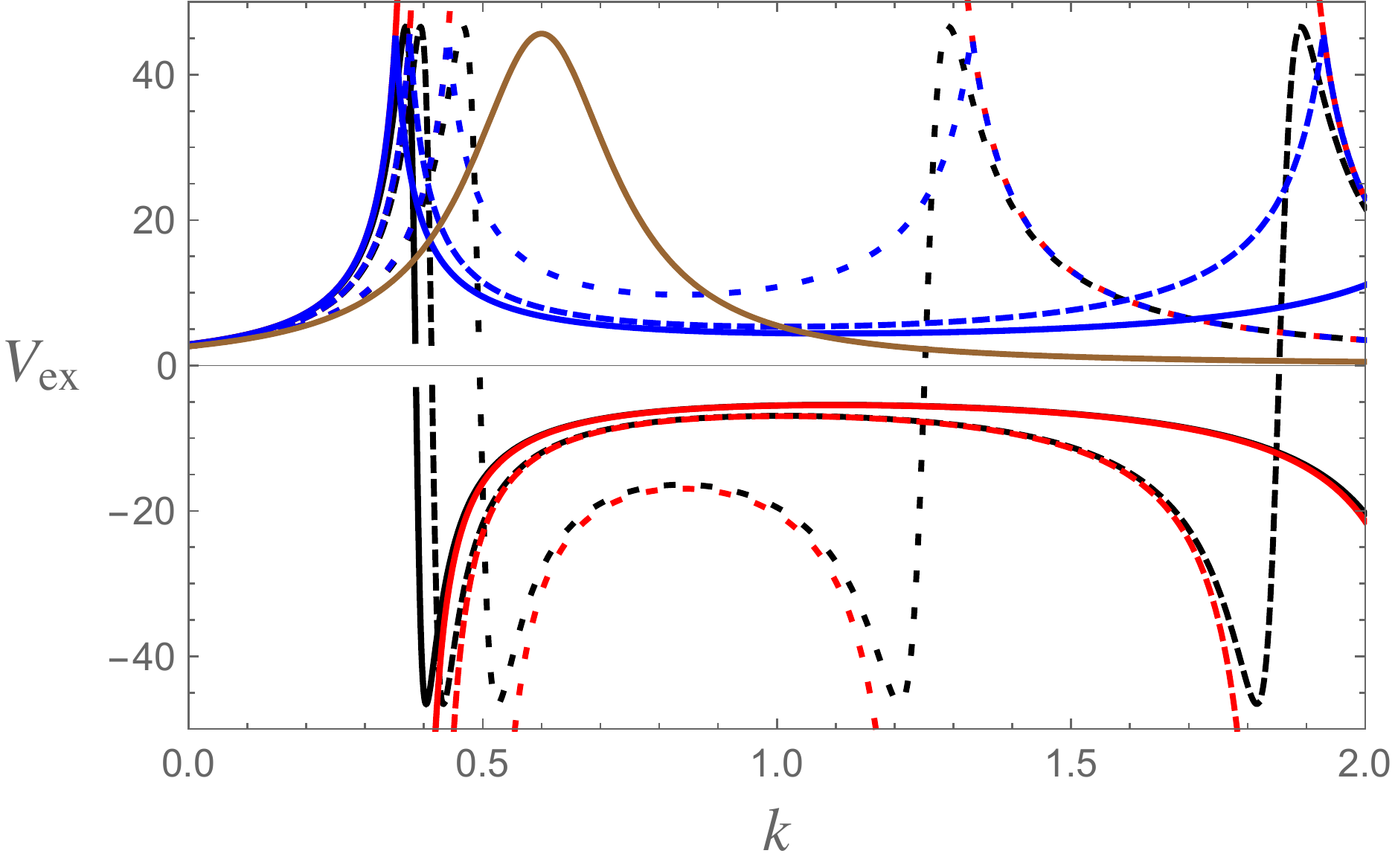}}
  \vspace{-2.22in}
\rightline{\includegraphics[height=2.2in]{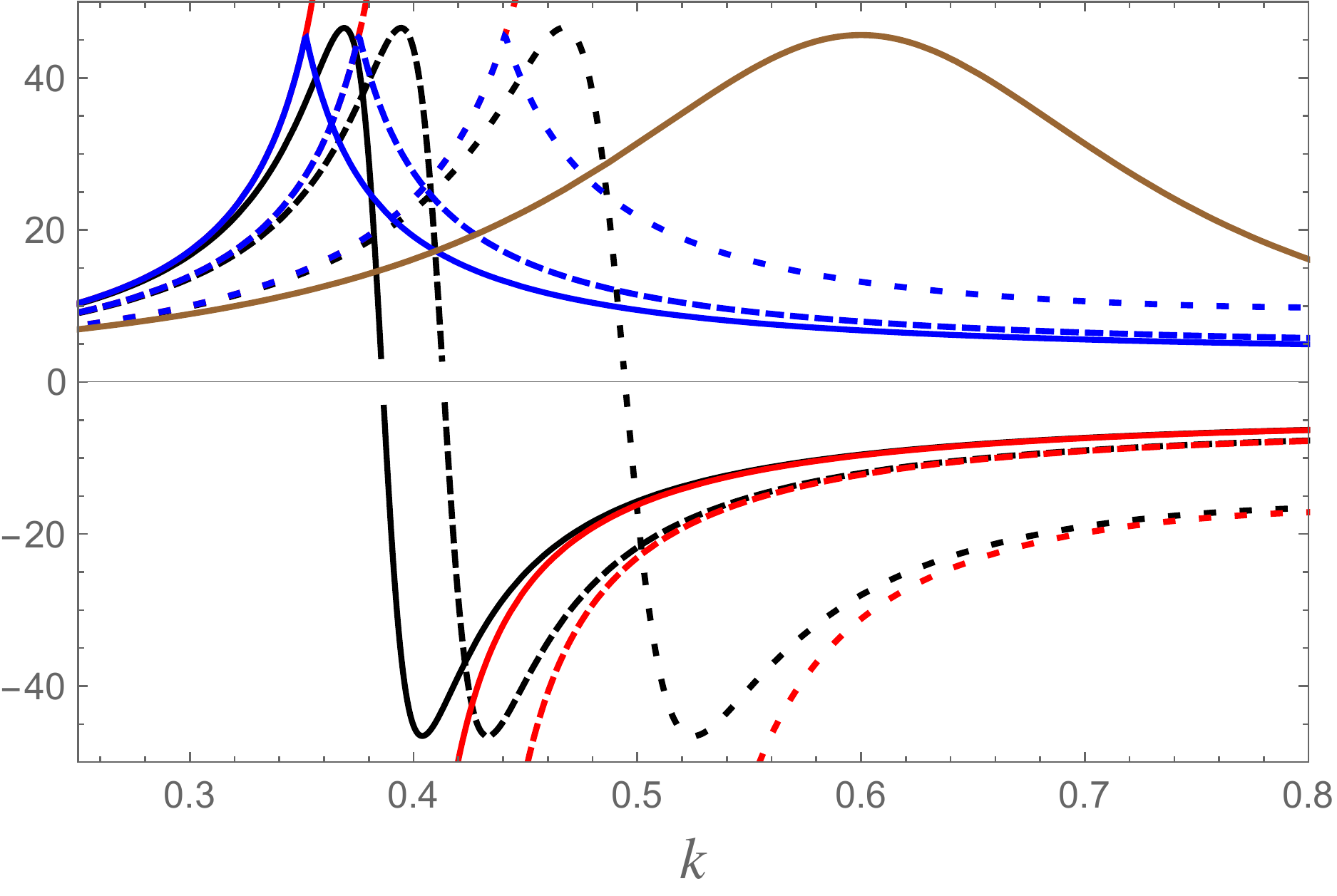}}
 \caption{{\footnotesize Left panel: lines for $z=-1$, $p=0.6$ showing the exchange kernel $V_{\rm ex}^{\rm D}$ (with $\lambda=0.7$) as a function of $k$ for $W=1.978$ (black), $W=2.01$ (long-dashed black) and $W=2.1$ (short-dashed black), the singular kernel $V_{\rm ex}^{\rm B}$ as a function of $k$ for $W=1.978$ (red), $W=2.01$ (long-dashed red) and $W=2.1$ (short-dashed red),  and the regularized kernel $V_{\rm ex}^{\rm C}$ as a function of $k$ for $W=1.978$ (blue), $W=2.01$ (long-dashed blue) and $W=2.1$ (short-dashed blue). The nonrelativistic limit, independent of energy,  is shown for comparison (brown). The dimensionless units are defined in the text.   Right panel: the same cases on an expanded scale around $k\sim 0.5$.}}
 \label{fig:exb}
\end{figure*}

\begin{figure*}
 \leftline{\includegraphics[height=2.2in]{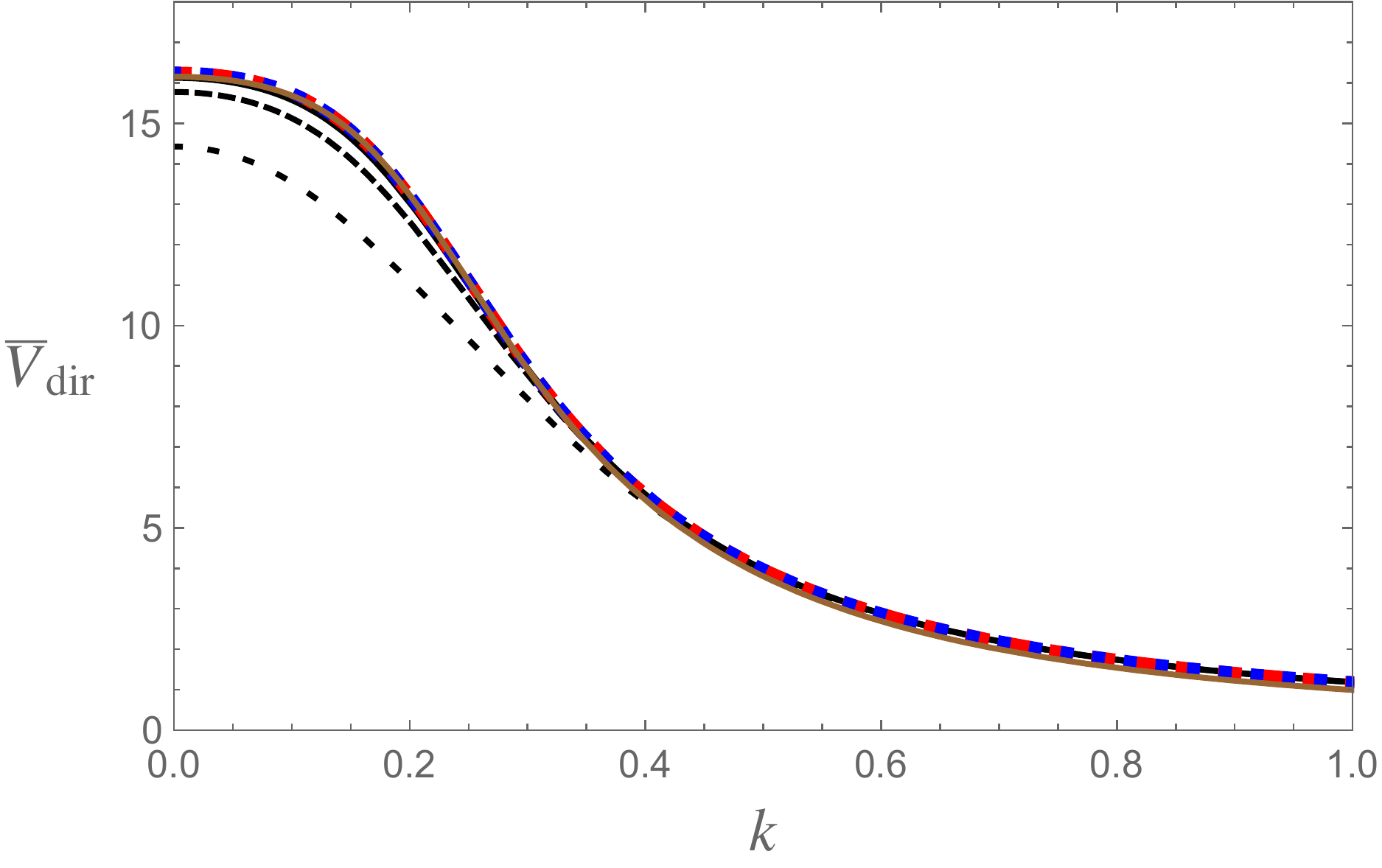}}
  \vspace{-2.23in}
\rightline{\includegraphics[height=2.2in]{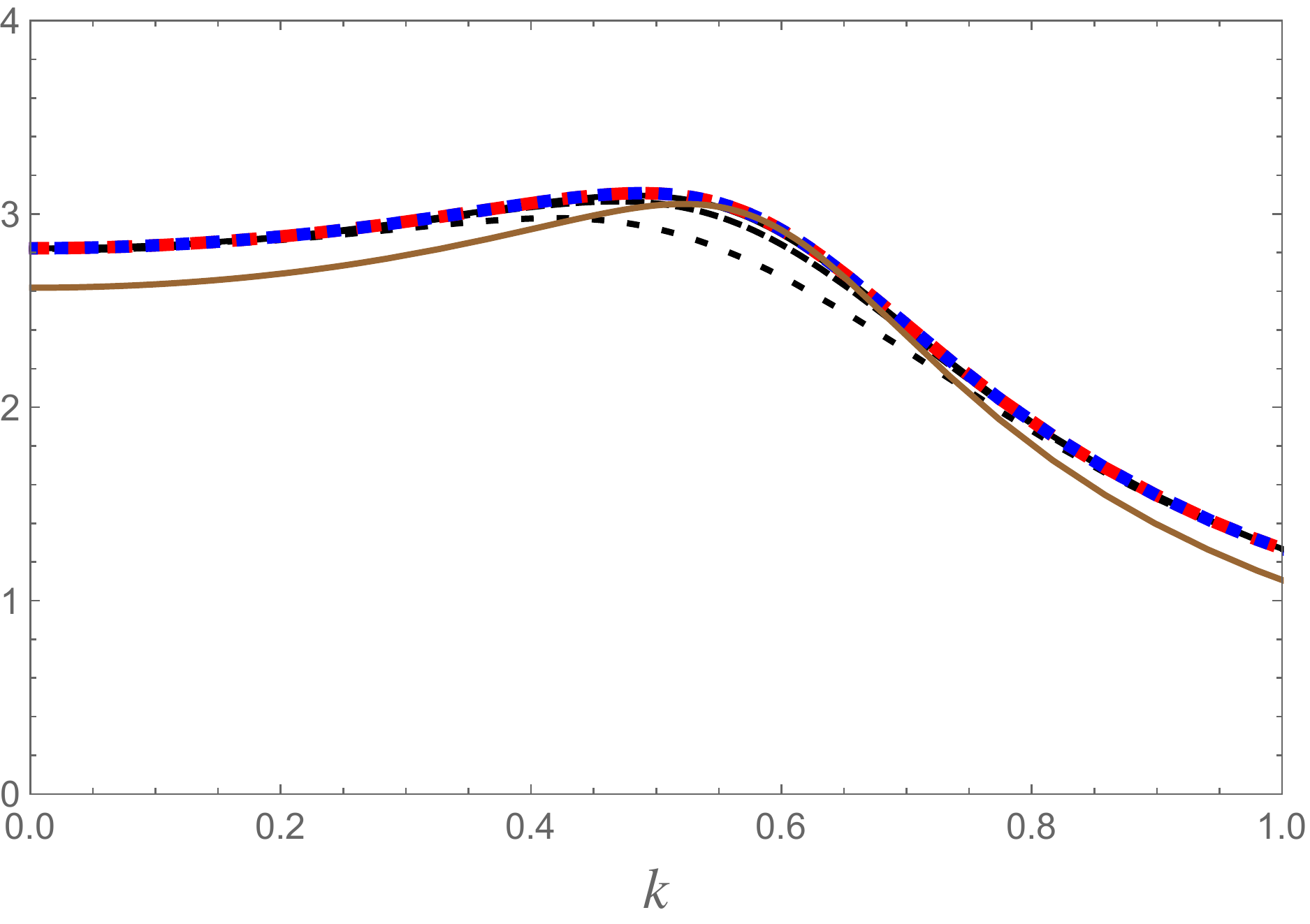}}
 \caption{{\footnotesize Left panel: lines for integrated kernels at $p=0.2$ showing the direct kernel $\bar{V}_{\rm dir}^{\rm D}$ as a function of $k$ for $\lambda=0.5$ (black), $0.7$ (long-dashed black) and $\lambda=1$ (short-dashed black).  The other cases, also for $p=0.2$, are the nonrelativistic limit (brown), $\bar{V}_{\rm dir}^{\rm B}$ (red dashed),  and $\bar{V}_{\rm dir}^{\rm C}$ (blue dashed).   The dimensionless units are defined in the text.  Right panel: the same cases but for $p=0.6$.}}
 \label{fig:dirInt}
\end{figure*}

\begin{figure*}
 \leftline{\includegraphics[height=2.2in]{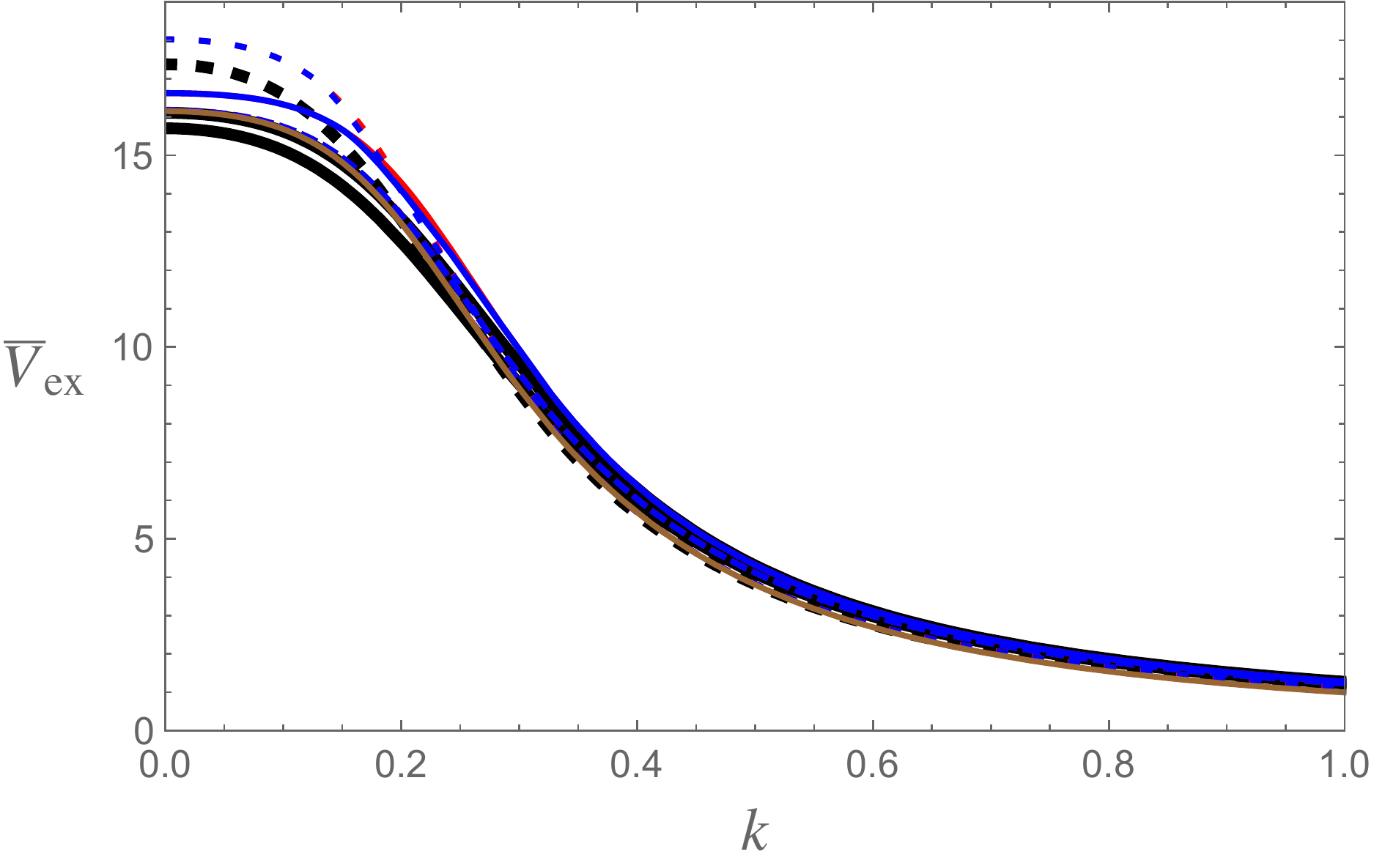}}
  \vspace{-2.23in}
\rightline{\includegraphics[height=2.2in]{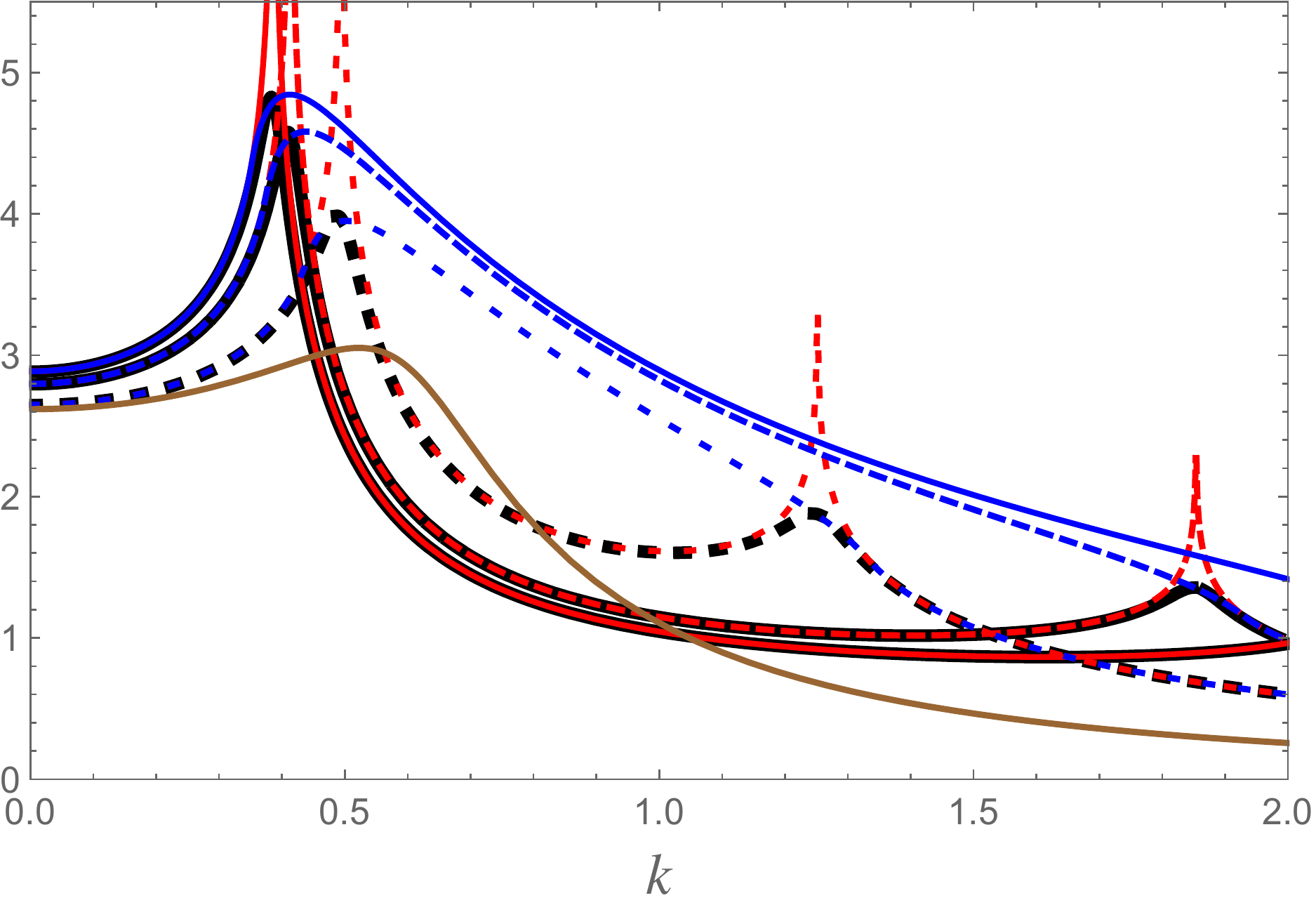}}
 \caption{{\footnotesize Left panel: lines for integrated kernels at $p=0.2$ showing the exchange kernel $\bar{V}_{\rm ex}^{\rm D}$ as a function of $k$ for $W=1.978$ (thick black), $W=2.01$ (thick long-dashed black) and $W=2.1$ (thick short-dashed black), the singular kernel $\bar{V}_{\rm ex}^{\rm B}$ as a function of $k$ for $W=1.978$ (red), $W=2.01$ (long-dashed red) and $W=2.1$ (short-dashed red), and the regularized kernel $\bar{V}_{\rm ex}^{\rm C}$ as a function of $k$ for $W=1.978$ (blue), $W=2.01$ (long-dashed blue) and $W=2.1$ (short-dashed blue). The nonrelativistic limit, independent of energy,  is shown for comparison (brown).  The dimensionless units are defined in the text.   Right panel: the same cases but for $p=0.6$.}}
 \label{fig:exInt}
\end{figure*}

\begin{figure*}
 \leftline{\includegraphics[height=2.2in]{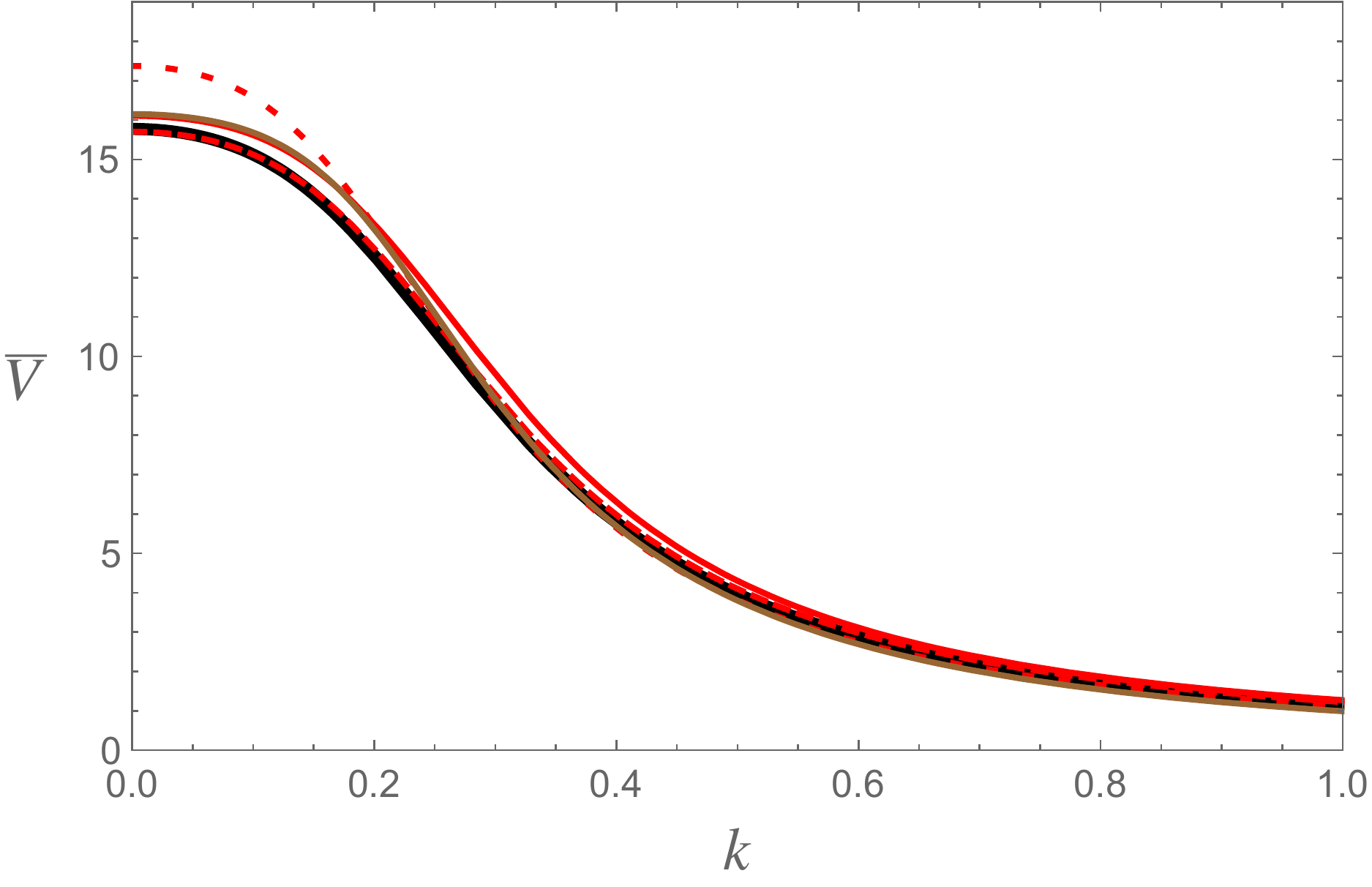}}
  \vspace{-2.23in}
\rightline{\includegraphics[height=2.2in]{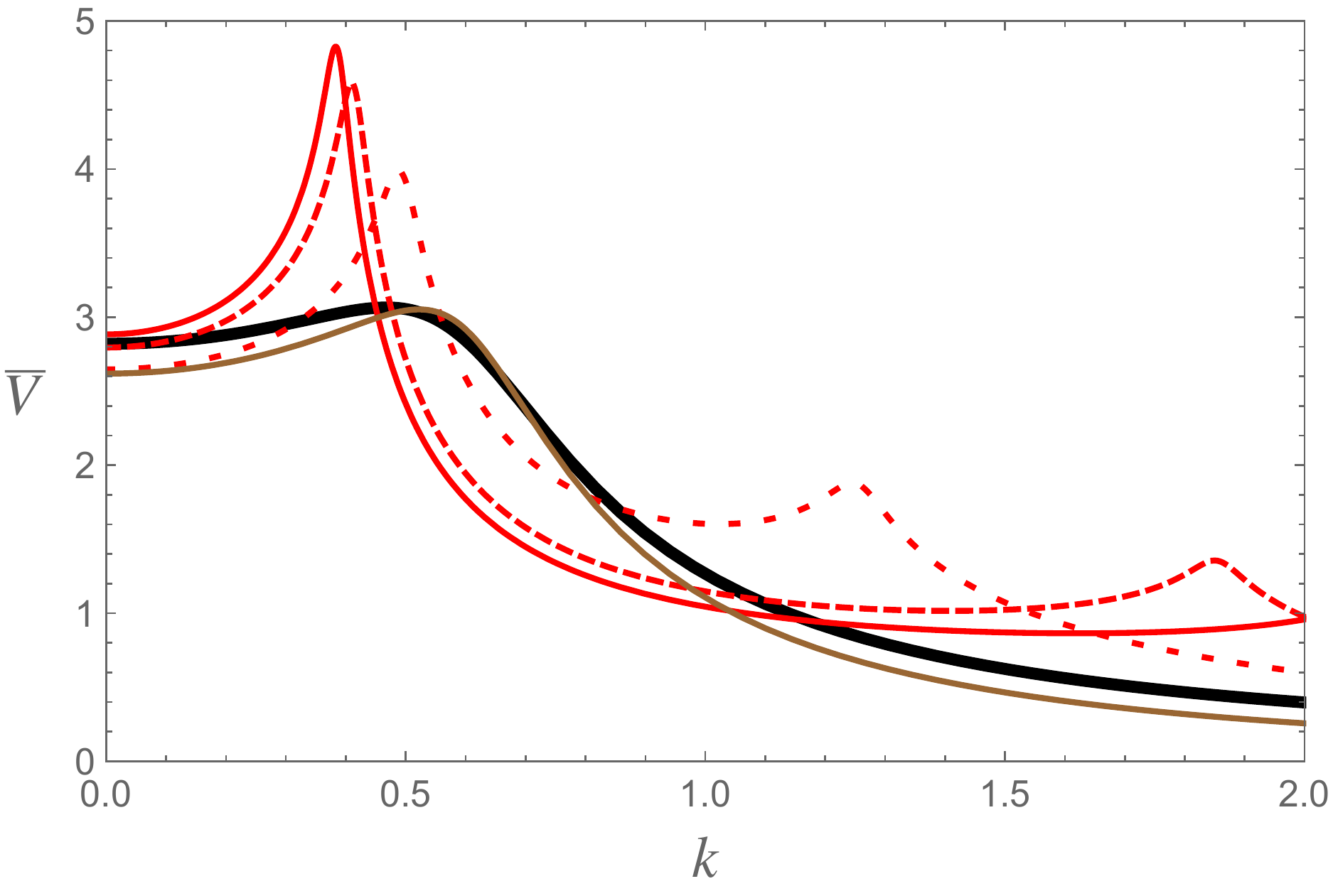}}
 \caption{{\footnotesize Left panel: lines at fixed $p=0.2$ as a function of $k$ for the energy independent  integrated kernel $\bar{V}_{\rm dir}^{\rm D}$ (thick black) and and the nonrelativistic limit (brown).  The energy dependent exchange kernel, $\bar{V}_{\rm ex}^{\rm D}$, is shown for $W=1.978$ (solid red), $W=2.01$ (long-dashed red) and $W=2.1$ (short-dashed red).  The dimensionless units are defined in the text.   Right panel: the same cases but for $p=0.6$.}}
 \label{fig:comp}
\end{figure*}


\subsection{Numerical implications}

Numerical comparisons of the three models B, C, and D are illustrated in this section.  
First, Figs.~\ref{fig:dir} -- \ref{fig:exb} show the functions
\bea
V_{\rm dir}&=&V(\hat k-\hat p)
\nonumber\\
{V}_{\rm ex}&=&V(P-\hat k-\hat p)
\eea
 for a fixed value of $z=\cos\theta$, where $z=1$ for the direct terms and $z=-1$ for the exchange terms (chosen so that the both terms agree with  the choice made in Fig.~\ref{fig:contour}).  All units are in the nucleon mass $M$ with $\mu/M=0.148$ and the $V$'s are divided by $-g^2\mu/M$.   These figures lead to the following conclusions: 
\begin{itemize}
\item The direct terms are all smooth. Model D depends significantly on the parameter $\lambda$.  Choosing $\lambda=0.7$ is a compromise between small $\lambda\simeq0.5$, which reproduces the other models, and large $\lambda\simeq 1$,  which strongly suppresses the peaks at $k\simeq p$. 
\item At small $p=0.2$, Fig.~\ref{fig:exa} shows that the exchange terms have no singularities, as shown already in Fig~\ref{fig:contour}, while Model C develops kinks that arise from the absolute value of $q^2$, and Model D (evaluated using $\lambda=0.7$) is quite similar to its direct counterpart.
\item At larger momenta $p$, Fig.~\ref{fig:exb} clearly shows the singularities that arise in Model B, how they are removed by Model C (at the expense of adding kinks), and how Model D (with $\lambda=0.7$) converts the singularities into smooth, but rapidly varying functions with an overall behavior similar to Model B.  (A larger $\lambda\simeq1$ will smooth out these variations significantly, but was not chosen for this discussion because it also significantly suppresses the direct term, as mentioned above.)
\item  For all models at large $p$, the $k$ dependence of the direct and exchange terms differs significantly in size and behavior.
\end{itemize}

The S-wave projections, $\bar V$, for each of these models are shown in Figs.~\ref{fig:dirInt} and \ref{fig:exInt} (in the same units as  Figs.~\ref{fig:dir} -- \ref{fig:exb}).  The direct integrals mirror the same behavior as the integrands shown in Fig.~\ref{fig:dir}, with the broad bump at $p\simeq k$ suppressed at higher $p$.  The choice $\lambda=0.7$ is still close to Model B, while $\lambda\simeq 1$ already shows significant deviations.  At large $p=0.6$, the exchange integrals for Model B show spikes and those for Model C are smooth but large.  Model D, with $\lambda=0.7$ follows Model B closely, eliminating the spikes.  Finally, Fig.~\ref{fig:comp} shows that the S-wave projections for the direct and exchange contributions are nearly identical at small $p$, but differ significantly at larger $p$, where the the direct term is still fairly close to the nonrelativistic result while the exchange term shows significant structure.

To study the convergence of the generalized ladder sum, and hence the extent that the OBE approximation dominates the solution, the fourth order diagrams are calculated next.

\begin{widetext}

\section{Fourth order diagrams}  \label{sec:three}

  \begin{figure*}[t]
\begin{center}
\includegraphics[width=6in]{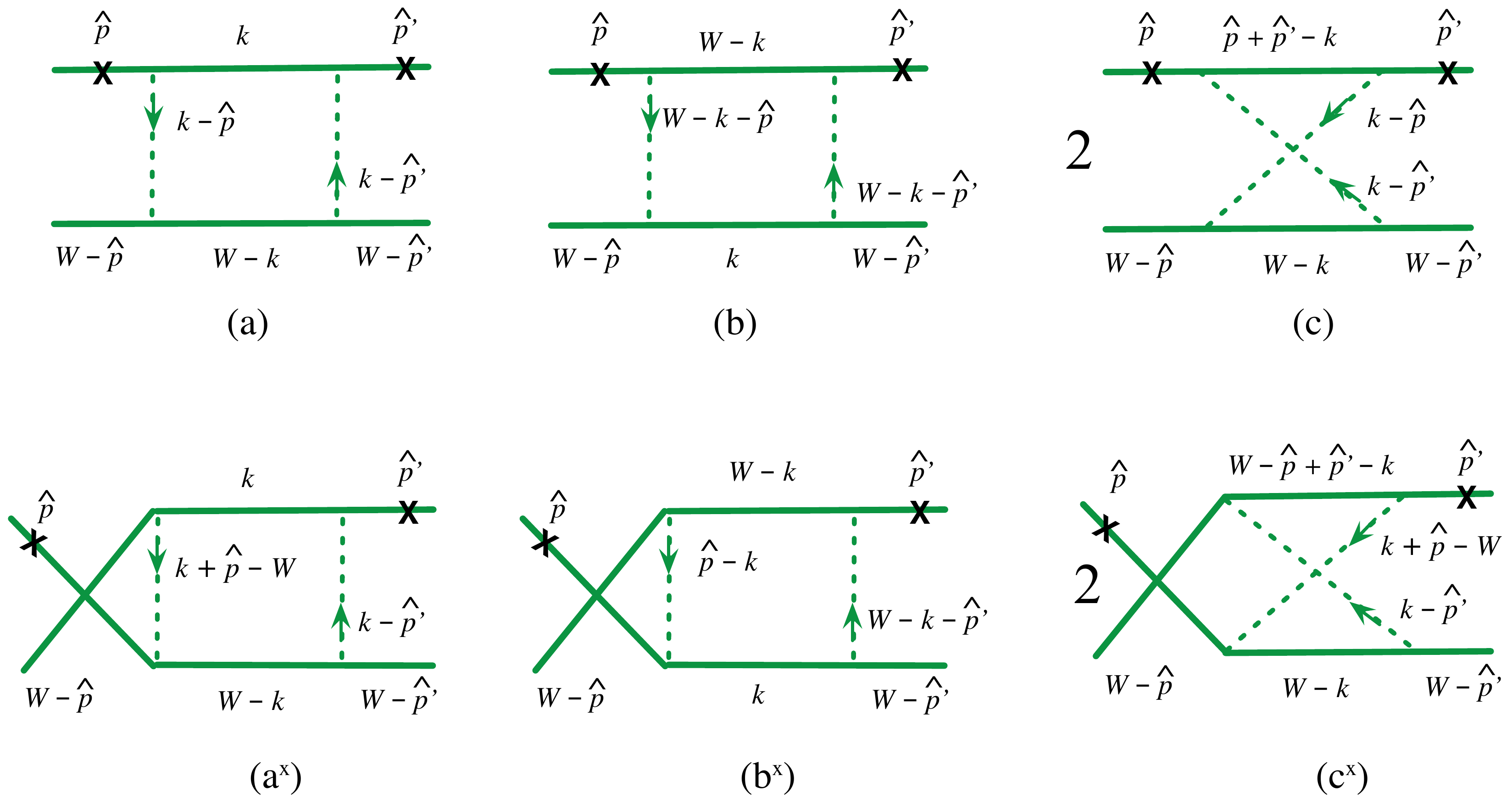}
\caption{Fourth order diagrams with particle 1 on shell in both the initial state and final states.   All diagrams are multiplied by one-quarter.   See the text for a full discussion.   }
\label{fig:fourthorder}
\end{center}
\end{figure*}

The generalized ladder diagrams for scattering to fourth order in CST are shown in Fig.~\ref{fig:fourthorder}.   For later comparison with the iterated OBE kernel (as suggested by Figs.~\ref{fig:2channel} and \ref{fig:2channel2}), it is convenient to symmetrize the box diagrams by including two direct diagrams, one with particle 1 having internal momentum $k$ (diagram a) and one with particle 2 having internal momentum $k$ (diagram b), and an exchange version of each (diagrams a$^x$ and b$^x$).  To avoid over counting, all of these box diagrams must be multiplied by one-quarter.  

The crossed box diagrams will be left unsymmetrized, and it is convenient to label their internal momenta so that the momentum of the mesons and particle 2 are identical to the box (a or a$^x$); therefore only the momenta of particle 1 is changed. Each of these diagrams  must be multiplied by one half  (the factor of one-quarter multiplying all diagrams in Fig.~\ref{fig:fourthorder} reduced to one-half by the factor of 2 in the figure). 

The sums of direct (top row) and exchange (second row) diagrams are then

\begin{subequations}
  \bea
 {M}_{\rm dir}^{4}&=&{M}_{\rm dir}^{4{\rm box}}+\delta_x {M}_{\rm dir}^{4{\rm xbox}}
\nonumber\\
&=&\frac{i}{2}\int\frac{d^4k}{(2\pi)^4} \Bigg\{\frac{V(\hat p-k)V(\hat p'-k)+V(P-\hat p-k) V(P-\hat p'-k)}{2\big[E_k^2-(W-k_0)^2\big](E_k^2-k_0^2-i\epsilon)}
\nonumber\\
&&\qquad\qquad\qquad+\frac{\delta_x V(\hat p-k)V(\hat p'-k)}{\big[E_k^2-(W-k_0)^2\big]\big[E_{k-p-p'}^2-(E_p+E_{p'}-k_0)^2\big]}\Bigg\}   
\label{eq:Vdir0}
 \\
 {M}_{\rm ex}^{4}&=&{M}_{\rm ex}^{4{\rm box}}+\delta_x {M}_{\rm ex}^{4{\rm xbox}}
 \nonumber\\
 &=&\frac{i}{2}\int\frac{d^4k}{(2\pi)^4} \Bigg\{\frac{V(P-\hat p-k)V(\hat p'-k)+V(\hat p-k)V(P-\hat p'-k)}{2\big[E_k^2-(W-k_0)^2\big](E_k^2-k_0^2-i\epsilon)} 
 \nonumber\\
 &&\qquad\qquad\qquad+\frac{\delta_x V(P-\hat p-k)V(\hat p'-k)}{\big[E_k^2-(W-k_0)^2\big]\big[E_{k+p-p'}^2-(W-E_p+E_{p'}-k_0)^2\big]}\Bigg\}
 \label{eq:Vex0}
 \eea
 \end{subequations}
 \end{widetext}
 where the first term in each curly bracket is  the symmetrized box 
and the second term is the crossed box, multiplied by $\delta_x=1$ (with momenta as shown in Fig.~\ref{fig:fourthorder}).  The $-i\epsilon$ prescription has been kept only for those poles that make leading contributions in the lower half $k_0$ complex plane (i.e. at $k_0=E_k$).

The total scattering to fourth order is  the sum of the two terms (\ref{eq:Vdir0}) and (\ref{eq:Vex0}) [the one-half in (\ref{eq:Vtotal}) is already been included in the definitions (\ref{eq:Vdir0}) and (\ref{eq:Vex0})]:
\bea
\bar{M}^4={M}_{\rm dir}^{4}+\eta{M}_{\rm ex}^{4}\, ,\label{eq:V4total}
\eea
and, by construction for $\eta=1$, is symmetric under interchange of the two baryons in the final state.  But the initial state must also be symmetric under the interchange $p'\leftrightarrow P-p'$, or 
\bea
\{E_{p'}, {\bf p}'\}\leftrightarrow \{W-E_{p'},-{\bf p}'\}\, . \label{eq:pprime}
\eea
 The box terms have this symmetry already. To prove the symmetry for the crossed box terms, first transform the exchange crossed box in (\ref{eq:Vex0}) using the substitution
\bea
k&\to& \hat p'-\hat p+k\, .\qquad
\label{eq:subs2}
\eea
This transforms the integrand of ${M}_{\rm ex}^{4{\rm xbox}}$ into
\bea
&&\frac{V(P-\hat p-k) V(\hat p'-k)}{\big[E_k^2-(W-k_0)^2\big]\big[E_{k+p-p'}^2-(W-E_p+E_{p'}-k_0)^2\big]} \to
\nonumber\\
&&\quad\frac{V(P-\hat p'-k) V(\hat p-k)}{\big[E_{k+p'-p}^2-(W+E_p-E_{p'}-k_0)^2\big]\big[E_{k}^2-(W-k_0)^2\big]}.
\nonumber\\ &&
\eea
In this form the two crossed box terms (\ref{eq:Vex0}) and (\ref{eq:Vdir0}) transform into each other under the exchange of the initial state particles, completing the demonstration.
 

\subsection{Iterated OBE and the subtracted box in CST }

The {\it irreducable\/} fourth order kernel, $\bar V^4$, is obtained from $\bar {M}^4$ by subtracting the second iteration of the OBE.  In the BS theory, the iteration is equal to the box, so the subtraction leaves only the crossed box.  However, in the CST, the iteration of the OBE is not equal to the box, and the remainder, the box minus the iteration of OBE, is the {\it subtracted\/} box.  

The iteration of the OBE kernel is defined by the first iteration of Eq.~(\ref{eq:2channel}), or diagrammatically in Fig.~\ref{fig:2channel2}.  When the direct part of this iteration is subtracted  from ${M}_{\rm dir}^{4{\rm box}}$, and the exchange part from ${M}_{\rm ex}^{4{\rm box}}$, the contributions from the pole at $k_0=E_k$ are zero (to get this result it was necessary to symmetrize the box, as we did).
Appendix \ref{app:A} gives the details of this subtraction.  
  

\subsection{Estimate of box contributions to $\bar{M}^4$ and $\bar V^4$}
\label{sec:m4andv4}

It is a straightforward matter to numerically evaluate the integrals (\ref{eq:Vdir0}) and (\ref{eq:Vex0}), but it is more instructive to obtain an approximate analytic result that displays their singularity structure.   To simplify the calculation, but still preserve the essential physics at low energy-momentum, consider the cases when $p,\,p'$ and $k$ are of ${\cal O}(\mu)$, and that $\mu\ll M$.  
This means that only  the 
three-momenta contained in the boson energies $\omega$ will be retained (i.e. $E_k\sim E_p\sim E_{p'}\to M$).  The total energy in the center of mass, $W$, will be written
\bea
W=2M+E_{\rm cm}\, ,
\eea
with $E_{\rm cm}$, the center of mass kinetic energy, also assumed to be of order $\mu$.  In scattering experiments, the cm energy is related to the lab energy by
\bea
E_{\rm cm}=2M\left(-1+\sqrt{1+\frac{E_{\rm lab}}{2M}}\right),
\eea
so a lab energy of 350 MeV$\simeq 2.5\mu$ (highest energy used in most $NN$ phase shift analyses) corresponds to a cm energy of about 1.2$\mu$, justifying this assumption.

A warning: these approximations mean that the only $k$ dependence kept for the evaluation the integral is in the boson energies; the $k$ dependence of $E_k$ is ignored, even when $k$ becomes very large, or when $E_k-M$ is very small.   The latter leads to a peculiar treatment of the propagator $G(\hat k,P)$, Eq.~(\ref{eq:G0}),  which becomes
\bea
G(\hat k,P)\simeq\frac{1}{2M(\frac{k^2}{M}-E_{\rm cm}-i\epsilon)}\to -\frac1{2ME_{\rm cm}}\, .
\eea
Without the factor of $k^2/M$, the details of the elastic cut structure will be lost; the estimates in this paper therefore differ in detail from those given in Ref.~\cite{Gross:1969rv}, but are sufficient to preserve the essential fact that the positive energy nucleon pole terms dominate at small $E_{\rm cm}$.  The large $k$ behavior of the integrands are also not treated accurately, but will   converge.  

With these assumptions, introducing $\kappa=k_0-M$ (so that $d^4k=d^3k dk_0=d^3kd\kappa$), the full box diagram is the sum of four terms 
\bea
&&{M}^{4 \rm box}_{ij}
\nonumber\\
&&\;\to \frac{i}{4}\int\frac{d^4k}{(2\pi)^4} \frac{(VV)_{ij}}{\big[M^2-(M+E_{\rm cm}-\kappa)^2-i\epsilon\big](-2M\kappa-i\epsilon)}
\nonumber\\
&&\;\to
i\int\frac{d^4k}{(2\pi)^4} \frac{(VV)_{ij}}{16M^2(-E_{\rm cm}+\kappa)(-\kappa-i\epsilon)} \label{eq:Mijbox}
\eea
where $i,j=\{1,2\}$ and

\begin{widetext}
 \bea
 (VV)_{11}=V(\hat p-k)V(\hat p'-k)&\to&\frac{g^4 M^2\mu^2}{(\omega_p^2-\kappa^2-i\epsilon)(\omega_{p'}^2-\kappa^2-i\epsilon)}
\nonumber \\
  (VV)_{12}(\hat p,\hat p')=V(\hat p-k)V(P-k-\hat p')&\to&\frac{g^4 M^2\mu^2}{(\omega_p^2-\kappa^2-i\epsilon)\big[\omega_{-p'}^2-(E_{\rm cm}-\kappa)^2-i\epsilon\big]}
\nonumber \\
   (VV)_{21}(\hat p,\hat p')=V(P-k-\hat p)V(\hat p'-k)&\to&(VV)_{12}(\hat p',\hat p)\nonumber \\
  (VV)_{22}=V(P-k-\hat p)V(P-k-\hat p')&\to&\frac{g^4 M^2\mu^2}{\big[\omega_{-p}^2-(E_{\rm cm}-\kappa)^2-i\epsilon\big]\big[\omega_{-p'}^2-(E_{\rm cm}-\kappa)^2-i\epsilon\big]}\, , \label{eq:VVs}
  \eea
where $F_b\to1$ in the last expression in each line.  
 Doing the $\kappa$ integrations gives
 \begin{subequations}
 \bea
 {M}^{4 \rm box}_{11}&=& \frac{g^4M\mu^2}{8}\! \!\int_k \Bigg[\underbrace{\frac{1}{E_{\rm cm}\omega_p^2\omega_{p'}^2}}_{\kappa=0\,{\rm pole}} +
 \frac{(\omega_p+\omega_{p'})(\omega_p+\omega_{p'}-E_{\rm cm})-\omega_{p}\omega_{p'}}{2\omega_p^2\omega_{p'}^2(\omega_p-E_{\rm cm})(\omega_{p'}-E_{\rm cm})(\omega_p+\omega_{p'})}\Bigg] \label{eq:M11box}
  \\
 {M}^{4 \rm box}_{12}(p,p')&=& {M}^{4 \rm box}_{21}(p',p)
 \nonumber\\
 &=&\frac{g^4M\mu^2}{8} \!\!\int_k \Bigg[ \underbrace{\frac{1}{E_{\rm cm}\,\omega_p^2(\omega_{-p'}^2-E_{\rm cm}^2)}}_{\kappa=0\,{\rm pole}}+
  \frac{(\omega_{p}+\omega_{-p'})\big[\omega_{p}^2+\omega_{-p'}^2 +\omega_{p}\omega_{-p'}+E_{\rm cm}(\omega_{-p'}-\omega_{p})\big]}
 {2\omega_{p}^2\omega_{-p'}^2(\omega_{-p'}+E_{\rm cm})(\omega_{p}-E_{\rm cm})\big[(\omega_{p}+\omega_{-p'})^2-E_{\rm cm}^2\big]}
 \Bigg]  
  \nonumber\\
  &=& \frac{g^4M\mu^2}{8}\!\int_k  \frac{2(\omega_{p}+\omega_{-p'})^2 \omega_{p}\,\omega_{-p'}-E_{\rm cm}(\omega_{p}+\omega_{-p'})^3-E_{\rm cm}\omega_{p}\,\omega_{-p'}+E_{\rm cm}^2(\omega_{p}^2+\omega_{-p'}^2)}
 {2E_{\rm cm}\, \omega_{p}^2 \omega_{-p'}^2(\omega_{p}-E_{\rm cm})(\omega_{-p'}-E_{\rm cm})\big[(\omega_{p}+\omega_{-p'})^2-E_{\rm cm}^2\big]}
 \label{eq:M12box}
\\
 {M}^{4 \rm box}_{22}&=& \frac{g^4M\mu^2}{8}\! \!\int_k \Bigg[\underbrace{\frac{1}{E_{\rm cm}(\omega_{-p}^2-E_{\rm cm}^2)(\omega_{-p'}^2-E_{\rm cm}^2)}}_{\kappa=0\,{\rm pole}} +
 \frac{(\omega_{-p}+\omega_{-p'})(\omega_{-p}+\omega_{-p'}+E_{\rm cm})-\omega_{-p}\omega_{-p'}}{2\omega_{-p}^2\omega_{-p'}^2(\omega_{-p}+E_{\rm cm})(\omega_{-p'}+E_{\rm cm})(\omega_{-p}+\omega_{-p'})} \Bigg]\qquad
 \nonumber\\
 &=& \frac{g^4M\mu^2}{8}\! \!\int_k 
 \frac{(\omega_{-p}+\omega_{-p'})\big[2\omega_{-p}\omega_{-p'}-E_{\rm cm}(\omega_{-p}+\omega_{-p'})+E_{\rm cm}^2\big]-E_{\rm cm}\omega_{-p}\omega_{-p'}}
 {2E_{\rm cm}\omega_{-p}^2\omega_{-p'}^2(\omega_{-p}-E_{\rm cm})(\omega_{-p'}-E_{\rm cm})(\omega_{-p}+\omega_{-p'})} \qquad  \label{eq:M22box}
 \eea
 \end{subequations} 
 
\end{widetext}
 where the contributions from the leading pole at $\kappa=0$ (originally $k_0=E_k$) have been identified, and the factor of $1/E_k\to 1/M$ in the integral over $k$.   As  shown in Appendix \ref{app:A},  {\it omitting\/} the results from the pole at $\kappa=0$ gives the irreducible kernel $V^{4{\rm box}}_{ij}$.

 \begin{table}[b]
\begin{minipage}{3.2 in}
\begin{ruledtabular}
\begin{tabular}{lcccc}
diagram &  \multicolumn{4}{c}{Singularities if $F_b=1$ (Model B)} \cr
& 2-boson & instability & scattering  & production  \cr 
&  $E_{\rm cm}^2\geq 4\mu^2$ & $E_{\rm cm}\leq\mu$ & $E_{\rm cm}=0$ & $E_{\rm cm}\geq \mu$ 
\\[0.05in]
\tableline
${M}^{4{\rm box}}_{11}$& & & x  & x \cr
${M}^{4{\rm box}}_{22}$& & & x  & x \cr
${M}^{4{\rm box}}_{12}, {M}^{4{\rm box}}_{21}$&x & & x  & x \cr
$V^{4{\rm box}}_{11}$ & &    &    & x \cr
$V^{4{\rm box}}_{22}$ & & x   &    & x \cr
$V^{4{\rm box}}_{12}, V^{4{\rm box}}_{21}$ & x& x &    & x \cr
\tableline
$V^{4{\rm xbox}}_{\rm dir}$ &  &  &    & x \cr
$V^{4{\rm xbox}}_{\rm ex}$ & x &  &    & x \cr
\end{tabular}
\end{ruledtabular}
\caption{Singularities in the amplitudes of the fourth order diagrams  discussed in this paper.} 
\label{tab:sig}
\end{minipage}
\end{table}

 To the accuracy we have been working, the singularities in the OBE exchange kernel, given in (\ref{eq:Wpm}), are at 
 \bea
 E_{\rm cm}=\pm\, \omega_{r} \, ,
 \eea
(where $r=\pm p$ or $\pm p'$), the plus sign corresponding to the physical production singularity and the minus sign to the unphysical instability singularity.  
Eqs.~(\ref{eq:M11box})--(\ref{eq:M12box}) show that the kernels $V^{4{\rm box}}_{ij}$ all have instability singularities (referred to simply as ``instability'' from now on), which are cancelled when the contributions from the leading nucleon ($\kappa=0$) poles are included.  The cancellation of the instability by another pole in the fourth order terms is the principal motivation for the subtraction (\ref{eq:OBED}).
 
 Note the presence of new singularities in ${M}^{4 \rm box}_{21}(p,p')$ at 
 \bea
 E_{\rm cm}=\pm\,(\omega_{-p}+\omega_{p'})\, .
 \eea
 These will be referred to collectively as ``2-boson'' singularities: one is a production singularity and one an instability.  That the singularity at  $E_{\rm cm}\leq -2\mu$ is due to an instability  can be seen by focusing on the lower nucleon line in Fig.~\ref{fig:fourthorder}.  The outgoing nucleon with four-momentum $\hat p$ will be unstable with respect to decay into the two bosons and the off-shell nucleon with momentum $P-\hat p'$ if
 \bea
 E_{-p}\geq W-E_{p'}+ \omega_{-p}+\omega_{p'}
 \eea
 which, with the approximations we are using, becomes
 \bea
 E_{\rm cm}\leq -(\omega_{-p}+\omega_{p'})\, ,\label{eq:2boson}
 \eea
 with the equality marking the boundary of the region of instability.

 The simplified discussion in Appendix \ref{app:B} shows how both of these 2-boson singularities are intrinsic to two-boson exchange (TBE), and introduce cuts into the TBE kernel.  To calculate the TBE kernel I will drop the imaginary parts from the cuts shown in Appendix \ref{app:B} and integrate over the remaining poles using the principal value prescription.  Possible improvements in this approach are beyond the scope of this paper, and are irrelevant to the to the final goal, which is to justify the neglect of the fourth-order kernel.  If the fourth-order kernel can be neglected, its detailed structure will play no role in subsequent calculations.

 Keeping track of the singularities in each of the amplitudes requires some care; Table \ref{tab:sig} provides a convenient summary.    I emphasize that the term ``instability'' will be understood to refer only to the one-boson instability of the type described by Eq.~(\ref{eq:instability}), and {\it not\/} any of the 2-boson singularities.
 
 In summary, the full box diagrams ${M}^{4{\rm box}}_{ij}$ have no instabilities, but have the scattering and production singularities.  The instability is only introduced when the dominant contribution from the pole at $\kappa=0$ is subtracted from ${M}^{4{\rm box}}_{ij}$, giving the irreducible kernels $V^{4{\rm box}}_{ij}$.  
 In the vicinity of the instabilities, which appear in $V^{4{\rm box}}_{12}, V^{4{\rm box}}_{21}, V^{4{\rm box}}_{22}$, these kernels cannot really be regarded as smaller that the iterated OBE, violating  the spirit of the cancellation theorem, and threatening the assumption that it is possible to ignore the fourth-order kernel.  Model D will solve this problem.  

 \subsection{Estimate of crossed box contributions to $V^4$}
 
The crossed box contributions are estimated starting from (\ref{eq:Vdir0}) and (\ref{eq:Vex0}) and using steps similar to those used for the evaluation of the box terms.  Noting that the crossed box is already irreducible, so that ${M}^{4{\rm xbox}}=V^{4{\rm xbox}}$, and dropping the factor of $\delta_x$ gives
\begin{widetext}
\begin{subequations}
\bea
 V_{\rm dir}^{4{\rm xbox}}&=&\,\frac{i}{2} \int\frac{d^4k}{(2\pi)^4}
 \frac{ V(\hat p-k)\, V(\hat p'-k)}{\big[M^2-(\hat p+\hat p'-k)^2\big]\big[M^2-(P-k)^2\big]} 
 \to\,i \int\frac{d^4k}{(2\pi)^4}
 \frac{V(\hat p-k)\, V(\hat p'-k)}{8M^2\kappa(\kappa-E_{\rm cm})} \label{eq:Vdirex}
  \\
 V_{\rm ex}^{4{\rm xbox}}&=&\,\frac{i}{2}\int\frac{d^4k}{(2\pi)^4} 
 \frac{V(P-\hat p-k)\, V(\hat p'-k)
 }{\big[M^2-(P-\hat p+\hat p'-k)^2\big]\big[M^2-(P-k)^2\big]}
 \to i\int\frac{d^4k}{(2\pi)^4} \frac{V(P-\hat p-k)\, V(\hat p'-k)} {8M^2(\kappa-E_{\rm cm})^2}
 \eea
 \end{subequations}
 where here none of the nucleon poles give leading contributions.  If $F_b=1$, the result becomes 
 \begin{subequations}
 \bea
 V^{\rm 4xbox}_{\rm dir}&=&i g^4 M^2\mu^2\!\!\int\frac{d^4k}{(2\pi)^4}
 \frac{ \big[(\omega_p^2-\kappa^2-i\epsilon)(\omega_{p'}^2-\kappa^2-i\epsilon)\big]^{-1}}{8M^2\kappa(\kappa-E_{\rm cm})}
 \to \frac{g^4M\mu^2}{8} \!\!\!\int_k \frac1{\omega_{p}^2-\omega_{p'}^2}
 \Bigg[\frac1{\omega_p^2(\omega_p-E_{\rm cm})}-\frac1{\omega_{p'}^2(\omega_{p'}-E_{\rm cm})}\Bigg]
 \nonumber\\
 &&=- \frac{g^4M\mu^2}{8}\int_k\frac{(\omega_p+\omega_{p'})(\omega_p+\omega_{p'}-E_{\rm cm})-\omega_p\omega_{p'}}{\omega_p^2\omega_{p'}^2 (\omega_p+\omega_{p'})(\omega_p-E_{\rm cm})(\omega_{p'}-E_{\rm cm})}\qquad\quad\label{eq:Vxdir0}
 \\
 V^{\rm 4xbox}_{\rm ex}&=& i g^4M^2\mu^2\!\!\int\frac{d^4k}{(2\pi)^4}
 \frac{ \big[(\omega_{-p}^2-(E_{\rm cm}-\kappa)^2-i\epsilon)(\omega_{p'}^2-\kappa^2-i\epsilon)\big]^{-1}}{8M^2(\kappa-E_{\rm cm})^2}
 \nonumber\\
 &\to&\frac{g^4M\mu^2}{8} \int_k \Bigg[\frac1{\omega_{-p}^3\big[(\omega_{-p}+E_{\rm cm})^2-\omega_{p'}^2\big]}+\frac1{\omega_{p'}(\omega_{p'}-E_{\rm cm})^2\big[(\omega_{p'}-E_{\rm cm})^2-\omega_{-p}^2\big]}\Bigg]
 \nonumber\\
 &=& -\frac{g^4M\mu^2}{8}\int_k  \frac{\omega_{p'}(\omega_{p'}+\omega_{-p}-E_{\rm cm})^2+\omega_{-p}^2(\omega_{-p}+\omega_{p'})}{\omega_{-p}^3\omega_{p'}(\omega_{p'}-E_{\rm cm})^2\big[(\omega_{-p}+\omega_{p'})^2-E_{\rm cm}^2\big]}\, . \label{eq:Vxex0}
 \eea
 \end{subequations}
 Neither of these has the elastic scattering or instability singularities, but both have production singularities and $V_{\rm ex}^{4{\rm xbox}}$ also has the 2-boson singularities discussed above. 
For later comparison, the fourth order kernel in the BS theory, in the limit when ${\bf p}$ and ${\bf p}'$  are very small, is
\bea
V^4_{\rm BS}= V_{\rm dir}^{4{\rm xbox}}+ V_{\rm ex}^{4{\rm xbox}}&\to& -\frac{g^4M\mu^2}{8} \int_k\Bigg\{\frac{3\omega-2E_{\rm cm}}{2\omega^4(\omega-E_{\rm cm})^2}+\frac{(2\omega-E_{\rm cm})^2+2\omega^2}{\omega^3(\omega-E_{\rm cm})^2(4\omega^2-E_{\rm cm}^2)}\Bigg\}
\nonumber\\ &=&
-\frac{g^4M\mu^2}{8}\int_k\frac{24\omega^3-16\omega^2 E_{\rm cm}-\omega E_{\rm cm}^2+2E_{\rm cm}^3 }{2\omega^4(\omega-E_{\rm cm})^2(4\omega^2-E_{\rm cm}^2)}=-\frac{g^4}{16}\int \frac{d^3k}{(2\pi)^3}\,\widetilde{v}^{4}_{\rm BS}\, . \label{eq:BS}
\eea

With this preparation, the total fourth order CST kernel will now be studied.

\end{widetext}

\section{The Cancellation Theorem} \label{sec:four}

As stated in the introduction, the cancellation theorem states that  the higher order kernels describing the  scattering of nonidentical scalar particles cancel when one of the nucleon masses approaches infinity, leaving the OBE ladders, summed by an integral equation with an OBE  kernel, to give the exact result for generalized sum of all ladders and crossed ladders. 

 For two identical particles with mass $M\to\infty$, it might appear that the theorem should follow from the case when they are not identical \cite{Gross:1969rv}, since the mass of the second particle did not affect the proof.  However, for nonidentical particles the OBE does {\it not\/} depend on the center of mass energy, while for identical particles the exchange term (required in order to build in the symmetry as discussed in Sec.~\ref{sec:two}) {\it does\/} depend on the center of mass energy and this is the essential difference that alters the results. 
 
 \subsection{Cancellations for small $E_{\rm cm}$}

 Begin the discussion by first considering the sum of the kernels $V_{11}^{4{\rm box}}$ and (because of the different symmetrization factors) one-half of $V_{\rm dir}^{4{\rm xbox}}$.  Using the general results (\ref{eq:Mijbox}) and (\ref{eq:Vdirex}) gives
\bea
\bar V^4_{11}&\equiv&V_{11}^{4{\rm box}}+\frac12V_{\rm dir}^{4{\rm xbox}}
\nonumber\\
&=&i\int\frac{d^4k}{(2\pi)^4} \frac{V(\hat p-k)V(\hat p'-k)}{16M^2(E_{\rm cm}-\kappa)}\Big[\frac1{\kappa}-\frac1{\kappa}\Big]\to 0\, ,\qquad
\eea 
{\it regardless\/} of the choice of the form factor $F_b$.  
These terms cancel only because the pole at $\kappa=0$ in (\ref{eq:Mijbox}) is included in ${M}_{11}^{4{\rm box}}$ but {\it not\/} in $V_{11}^{4{\rm box}}$; i.e. the result of dropping the $\kappa=0$ pole contributions from (\ref{eq:M11box}) cancels one-half of (\ref{eq:Vxdir0}).  The conclusion is that the cancellation theorem for $\bar V^4_{11}$ is proved to fourth order for any choice of form factor $F_b$ and for ``large'' $E_{\rm cm}\sim \omega$. 

\begin{widetext}

Collecting all of the remaining terms together gives
\bea
V^4_{\rm CST}&\equiv&V_{\rm ex}^{4{\rm xbox}}+V_{12}^{4{\rm box}}+V_{21}^{4{\rm box}}+ V_{22}^{4{\rm box}}+\frac12 V_{\rm dir}^{4{\rm xbox}}
\nonumber\\
&=&-i\int \frac{d^4k}{(2\pi)^4}\Bigg\{-\frac{(VV)_{21}}{8M^2(\kappa-E_{\rm cm})^2}
+\frac{(VV)_{12}+(VV)_{21}+(VV)_{22}-(VV)_{11}}{16M^2\kappa(\kappa-E_{\rm cm})}\Bigg\} 
\label{eq:Vrem}
\eea
where the $(VV)_{ij}$ were defined in (\ref{eq:VVs}).  If $E_{\rm cm}\sim p^2/M\to0$, then $P\to P_0= \{2M,{\bf 0}\}$, and
\bea
\lim_{E_{\rm cm}\to0}V^4_{\rm CST}&=&-i\int \frac{d^4k}{(2\pi)^4}\frac{1}{16M^2\kappa^2}\Big[V(\hat p-k) +V(P_0-\hat p-k)\Big]
\Big[V(P_0-\hat p'-k)-V(\hat p'-k)\Big]\to0\, ,
\eea
\end{widetext}
because, for both $\hat p$ and $\hat p'$, $(\hat p-k)^2\to\kappa^2-({\bf p}-{\bf k})^2$ and $(P-\hat p-k)^2\to \kappa^2-({\bf p}+{\bf k})^2$, and hence, for any kernel $V$ dependent only on $q^2$, the integrand is  odd in ${\bf k}$ and integrates to zero (even before the $\kappa$ integration is done).   The cancellation theorem is proved if the center of mass energy is small.

\subsection{Violations when $E_{\rm cm}$ is not small}

When $E_{\rm cm}\sim \mu$, the remainder term (\ref{eq:Vrem}) is no longer zero and the cancellation theorem is already violated at fourth order.  To simplify the study of these violations, and to see how the form factor that accompanies Model D modifies the results, the discussion will be limited to the special case that ${\bf p}$ and ${\bf p}'$ are very small.  In this case all of the boson energies are equal, and will be denoted by $\omega=\sqrt{\mu^2+k^2}$.  Furthermore, in this limit, $(VV)_{12}=(VV)_{21}$, and $(VV)_{11}(\kappa)=(VV)_{22}(\kappa-E_{\rm cm})$, so $\kappa\to \kappa'+E_{\rm cm}$ shifts the $(VV)_{22}$ terms into the $(VV)_{11}$ form, so many terms may be combined, giving two distinct contributions
\begin{widetext}
\begin{subequations}
\bea
V^{4{\rm B}}_{21}\equiv V_{\rm ex}^{4{\rm xbox}}+V_{12}^{4{\rm box}}+V_{21}^{4{\rm box}}&\to&
i\int \frac{d^4k}{(2\pi)^48M^2}\frac{(VV)_{21}}{(\kappa-E_{\rm cm})}\bigg[\frac1{\kappa-E_{\rm cm}}-\frac1{\kappa}\bigg]=i\int \frac{d^4k}{(2\pi)^48M^2}\frac{E_{\rm cm}\,(VV)_{21}}{\kappa(\kappa-E_{\rm cm})^2}
\nonumber\\
&=&-\frac{g^4 M\mu^2}{8}\int_k\frac{E_{\rm cm}(8\omega^2-3\omega E_{\rm cm}+E_{\rm cm}^2)}{\omega^3(\omega-E_{\rm cm})^2(\omega+E_{\rm cm})(4\omega^2-E_{\rm cm}^2)} \label{eq:combos1}
\\
V^{4{\rm B}}_{11}\equiv V_{22}^{4{\rm box}}+\frac12 V_{\rm dir}^{4{\rm xbox}}&\to&i\int \frac{d^4k}{(2\pi)^4 16M^2}\frac{(VV)_{11}}{\kappa}\bigg[\frac1{\kappa-E_{\rm cm}}-\frac1{\kappa+E_{\rm cm}}\bigg]=i\int \frac{d^4k}{(2\pi)^4 8M^2}\frac{E_{\rm cm}(VV)_{11}}{\kappa(\kappa^2-E_{\rm cm}^2)}
\nonumber\\
&=&-\frac{g^4 M\mu^2}{8}\int_k\frac{E_{\rm cm}(2\omega^2-E_{\rm cm}^2)}{\omega^4(\omega^2-E_{\rm cm}^2)^2}\label{eq:combos2}
\eea
\end{subequations}
Hence the total CST fourth order kernel is
\bea
{V}^4_{\rm B}&\to&
-\frac{g^4 M\mu^2}{8}\int_k\frac{E_{\rm cm}(16\omega^4+5E_{\rm cm}\omega^3 - 8 E_{\rm cm}^2 \omega^2  + E_{\rm cm}^3 \omega + E_{\rm cm}^4)}{\omega^4(\omega^2-E_{\rm cm}^2)^2(4\omega^2-E_{\rm cm}^2)}=-\frac{g^4}{16}\int \frac{d^3k}{(2\pi)^3}\,\widetilde{v}^{4}_{\rm B}
\, .
  \label{eq:cancellation2}
\eea
\end{widetext}
These results clearly display the cancellation as $E_{\rm cm}\to0$, and    
the singularities from OBE instability and two-boson exchange.

The form factor that accompanies Model D, will modify these results.  
While the form factor is real along the real $q_0$ axis, it has four poles in the complex $q_0$ plane with a location depending on $q^2$.  These poles are at
\bea
q_0&=&\epsilon_1\sqrt{\omega^2+i\epsilon_2  \lambda_\mu^2 -i\epsilon}
\nonumber\\
&=&\epsilon_1 \left[R_\omega+i\epsilon_2  I_\omega-i\epsilon'\right] 
\eea
where 
$\omega$ is the appropriate boson energy for the particular $q$, the phases $\epsilon_1$ and $\epsilon_2$ are independent and equal to $\pm1$,  and
\bea
r^2&=&\sqrt{\omega^4+\lambda_\mu^4}
\nonumber\\
R_\omega&=&\frac{\sqrt{r^2+\omega^2}}{\sqrt{2}}
\nonumber\\
 I_\omega&=&\frac{\lambda_\mu^2}{2 R_\omega}\qquad
\epsilon'\sim \frac{\epsilon}{2 R_\omega}\, . \label{eq:subsdefs}
\eea
Hence the form factor can be written (ignoring the $i\epsilon$ in the numerator)
\bea
F_b(q^2)&&=\frac{(\omega^2-q_0^2)^2}{D_b} ,\quad
\eea
where the denominator factors into four poles, numbered for later reference:
\bea
D_b&&=\underbrace{(R_\omega+i\,I_\omega -q_0-i\epsilon')}_{1}\underbrace{(R_\omega-i\,I_\omega -q_0-i\epsilon')}_{2}
\nonumber\\
&&\qquad\times \underbrace{(R_\omega+i\,I_\omega +q_0-i\epsilon')}_{3}\underbrace{(R_\omega-i\,I_\omega +q_0-i\epsilon')}_{4}
\nonumber\\
&&=(R_\omega+i\,I_\omega -q_0-i\epsilon')(R_\omega-i\,I_\omega -q_0-i\epsilon')
\nonumber\\
&&\qquad\qquad\times \big[I_\omega^2+(R_\omega+q_0)^2\big]
\nonumber\\
&&=\lambda_\mu^4+(\omega^2-q_0^2-i\epsilon)^2\, .\qquad
\label{eq:Db}
\eea
  
 \begin{figure}[b]
 \centering
 \includegraphics[width=3in]{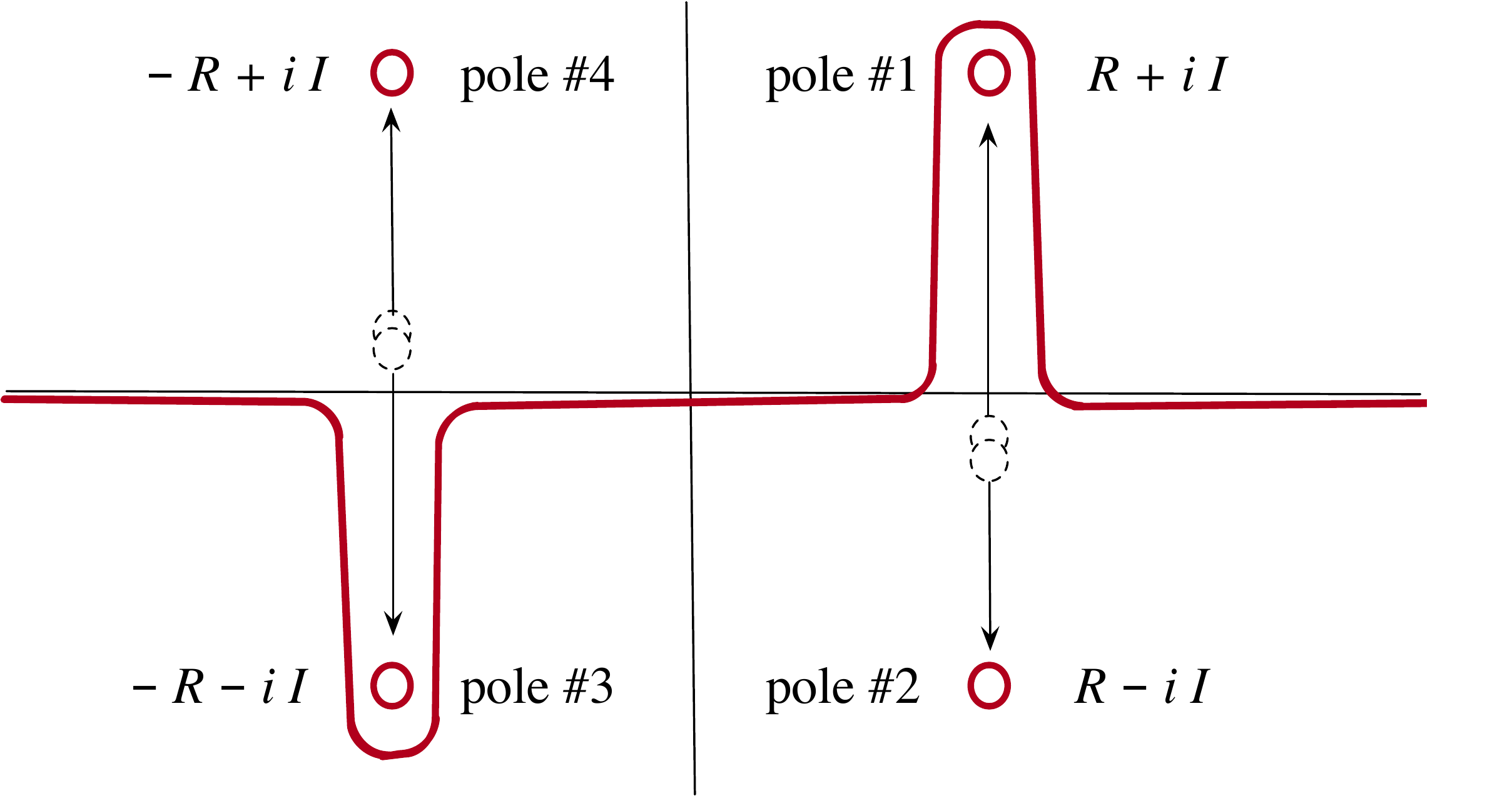}
 \caption{Figure showing the migration of the four poles identified in Eq.~(\ref{eq:Db}) in the complex $q_0$ plane.  Two double poles for $I=0$ are shown  as overlapping dashed circles; as $I$ increases they migrate to the small red circles, and the contour (red line) is deformed to avoid them. }
 \label{fig:contour2}
\end{figure}


In evaluating these contributions, great care must be taken with contour integrations.   For consistency with the case when $\lambda = 0$, the physical sheet is defined when $\lambda_\mu^2<\epsilon$; for larger $\lambda_\mu$ the expressions  must be analytically continued from this physical sheet.  Therefore, as $\lambda_\mu$ (or $I_\omega$) increases from zero, pole 2 remains in the lower half plane and pole 4 remains in the upper half plane.  However, pole 1(3) starts in the lower(upper) half plane but crosses the real axis into the upper(lower) half plane.    
To remain on the same sheet, the contour must be deformed around poles 1 and 3 as shown in Fig.~\ref{fig:contour2}.  Since the contour is being closed in the lower half plane, the physical sheet is defined by the contributions from poles 1 and 2, even though, for large $\lambda_\mu$, it would appear at first sight that pole 3 should be included and not pole 1 (this problem was first encountered over 20 years ago \cite{Ramalho:1998xg}).

The contributions from the poles 1 and 2 from each meson exchange  fix $\kappa$ at two values,
\bea
\kappa_\omega^\pm=R_{\omega}\pm i\,I_{\omega}=\frac1{2R_{\omega}}(2R_{\omega}^2\pm i \lambda_\mu^2)\, .
\eea
These are complex conjugates, so the contribution from each exchange is twice the real part of the contribution from pole 1. Note the convenient simplification 
\bea
I_\omega\big[I_\omega^2+(R_\omega+\kappa_\omega^+)^2\big]=4I_\omega R_\omega\kappa_\omega^+ =2\lambda_\mu^2  \kappa_\omega^+\, . \qquad
\eea
Hence, when ${\bf p}$ and ${\bf p}'$ are very small, the Model D integral in (\ref{eq:combos1}) for $(VV)_{21}$  generalizes to
\begin{widetext}
\bea
&&V^{4{\rm D}}_{21}=i\int\frac{d^4 k}{(2\pi)^4} \frac{E_{\rm cm}(V^{\rm D}V^{\rm D})_{21}}{8M^2\kappa(\kappa-E_{\rm cm})^2}
\nonumber\\
&&\quad=\frac{g^4 M\mu^2}4 \Re\int_k  \frac{E_{\rm cm}\big[\omega^2-(\kappa_\omega^+)^2\big]}{2i \lambda_\mu^2 (\kappa_\omega^+)^2}\left\{
\frac{\big[\omega^2-(\kappa_\omega^+-E_{\rm cm})^2\big] (\kappa_\omega^+-E_{\rm cm})^{-2}}
{\Big\{\lambda_\mu^4+\big[\omega^2-(\kappa_\omega^+-E_{\rm cm})^2\big]^2\Big\}}
+ \frac{\big[\omega^2-(\kappa_\omega^++E_{\rm cm})^2
\big] (\kappa_\omega^++E_{\rm cm})^{-1}}
{\kappa_\omega^+
\Big\{\lambda_\mu^4+\big[\omega^2-(\kappa_\omega^++E_{\rm cm})^2\big]^2\Big\}}\right\}
\nonumber\\
&&\quad= -\frac{g^4}{16}\int \frac{d^3k}{(2\pi)^3}\,\widetilde{v}^{4{\rm D}}_{21}\, . \label{eq:V21D}
\eea
Note that the $\lambda_\mu\to0$ limit of this integral is easily obtained from the limits $R_\omega\to \omega+{\cal O}(\lambda_\mu^4)$, $I_\omega\to \lambda_\mu^2/(2\omega)+{\cal O}(\lambda_\mu^4)$ and $\omega^2-(\kappa^+_\omega)^2\to - i\lambda_\mu^2+{\cal O}(\lambda_\mu^4)$, so that
\bea
\lim_{\lambda_\mu\to0} V^{4{\rm D}}_{21}&=&-\frac{g^4 M\mu^2 E_{\rm cm}}{4}\int_k\Bigg\{\frac{1}{2\omega^2(\omega-E_{\rm cm})^2(2\omega-E_{\rm cm})}-\frac{1}{2\omega^3(\omega+E_{\rm cm})(2\omega+E_{\rm cm})}\Bigg\}
\nonumber\\
&=&-\frac{g^2 M\mu^2}{8}\int_k\frac{E_{\rm cm}(8\omega^2-3\omega E_{\rm cm}+E_{\rm cm}^2)}{\omega^3(\omega-E_{\rm cm})^2(\omega+E_{\rm cm})(4\omega^2-E_{\rm cm}^2)}
\eea
in agreement with (\ref{eq:combos1}).

\begin{figure*}
 \leftline{\includegraphics[height=2.2in]{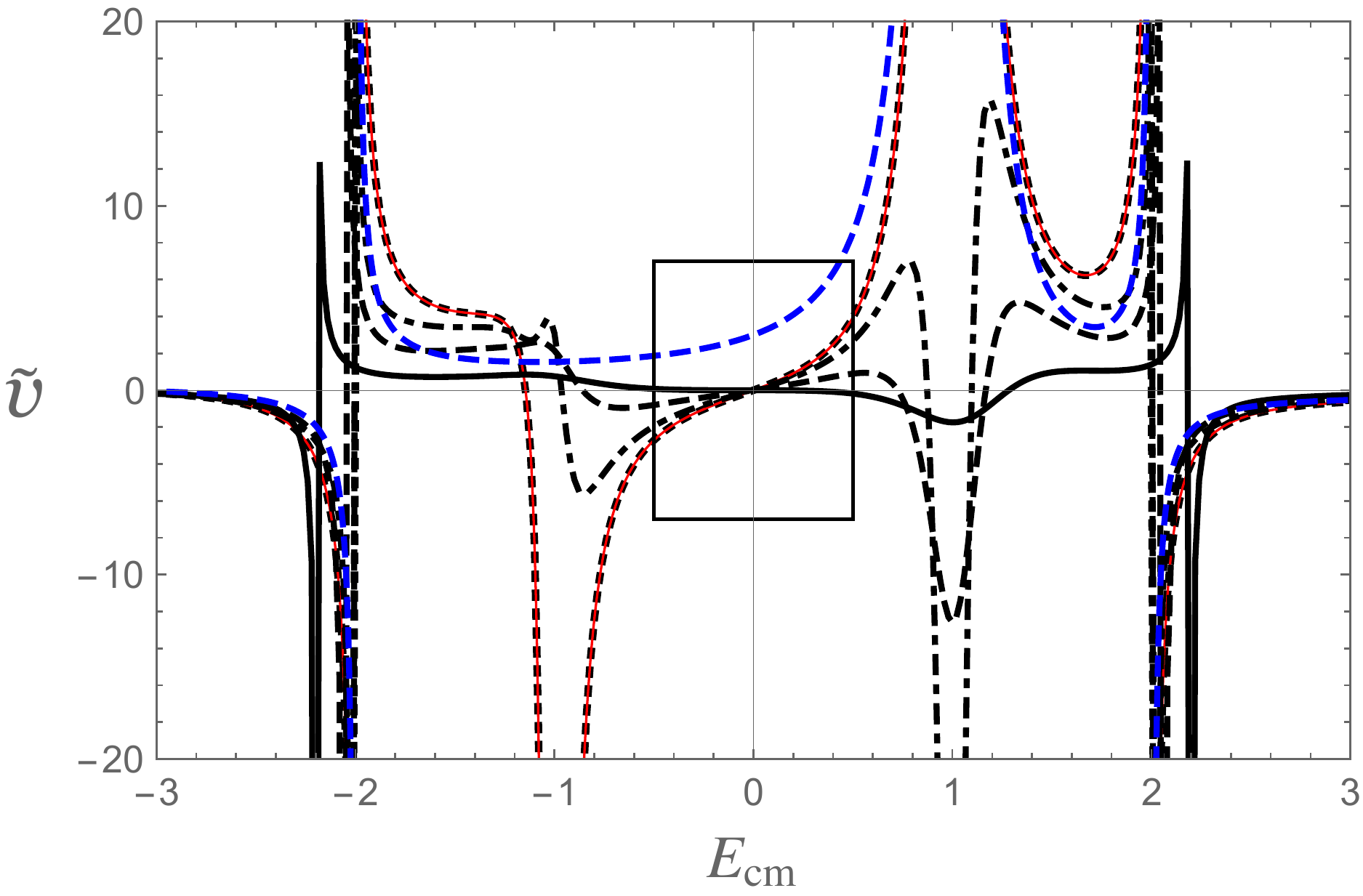}}
  \vspace{-2.2in}
\rightline{\includegraphics[height=2.2in]{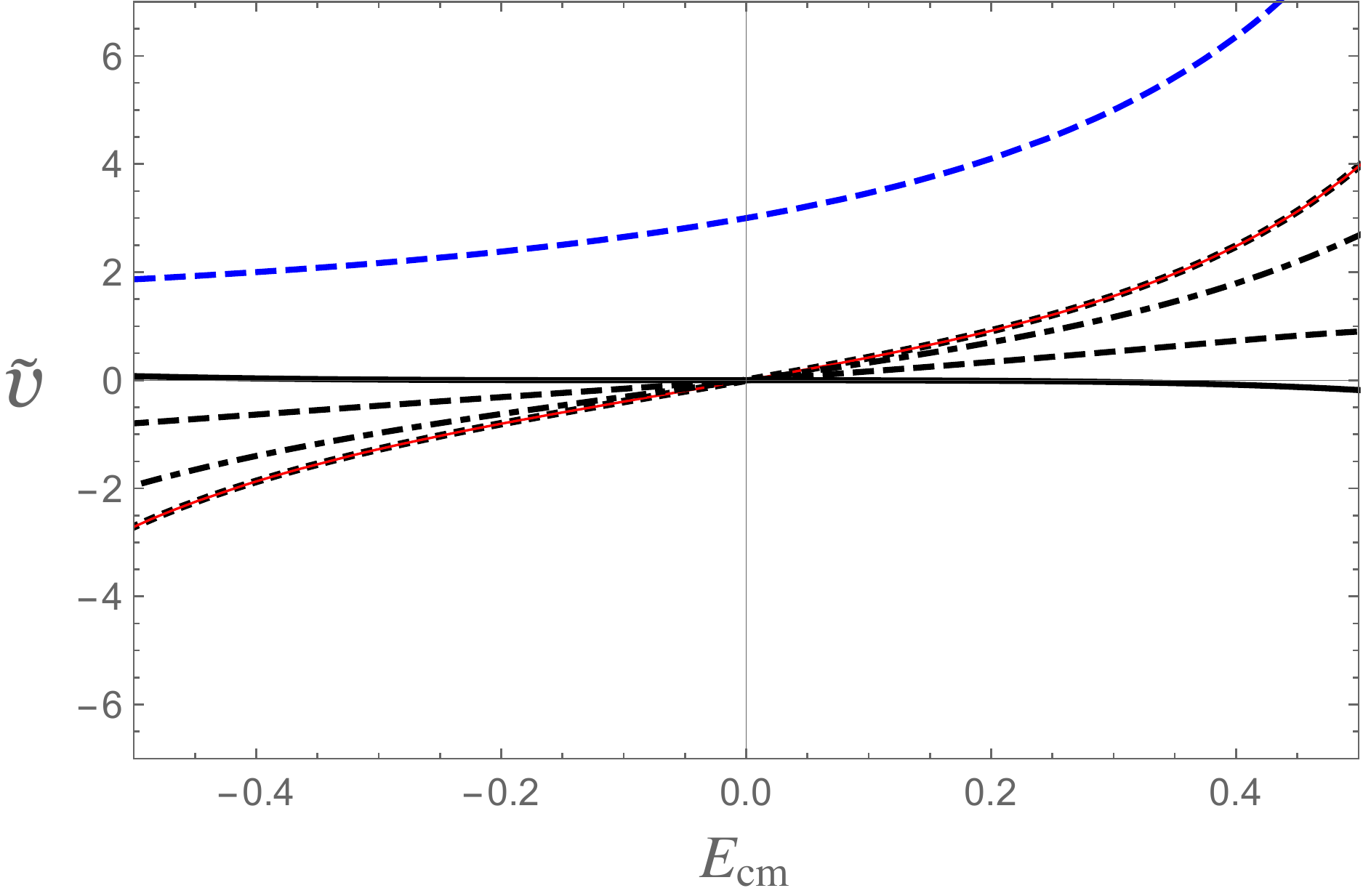}}
 \caption{{\footnotesize Lines showing the integrands 
 $\widetilde{v}^{4}_{\rm BS}$ (blue dashed), $\widetilde{v}^{4}_{\rm B}$ (thin red) and $\widetilde{v}^{4}_{\rm D}$ for $\lambda=0.1$ (black dotted line overlapping the thin red line), 0.5 (black dot-dashed), 0.7 (black dashed), and 1 (black solid), all  as a function of $E_{\rm cm}$ for the special case when $\mu=1$ and  $p=p'=k=0$.  Left panel: expanded scale.  Right  panel: enlarged picture of the region in the square box in the left panel.  Units for $E_{\rm cm}$ are the boson mass $\mu$.}}
 \label{fig:tildeV}
\end{figure*}

\begin{figure*}
 \leftline{\includegraphics[height=2.3in]{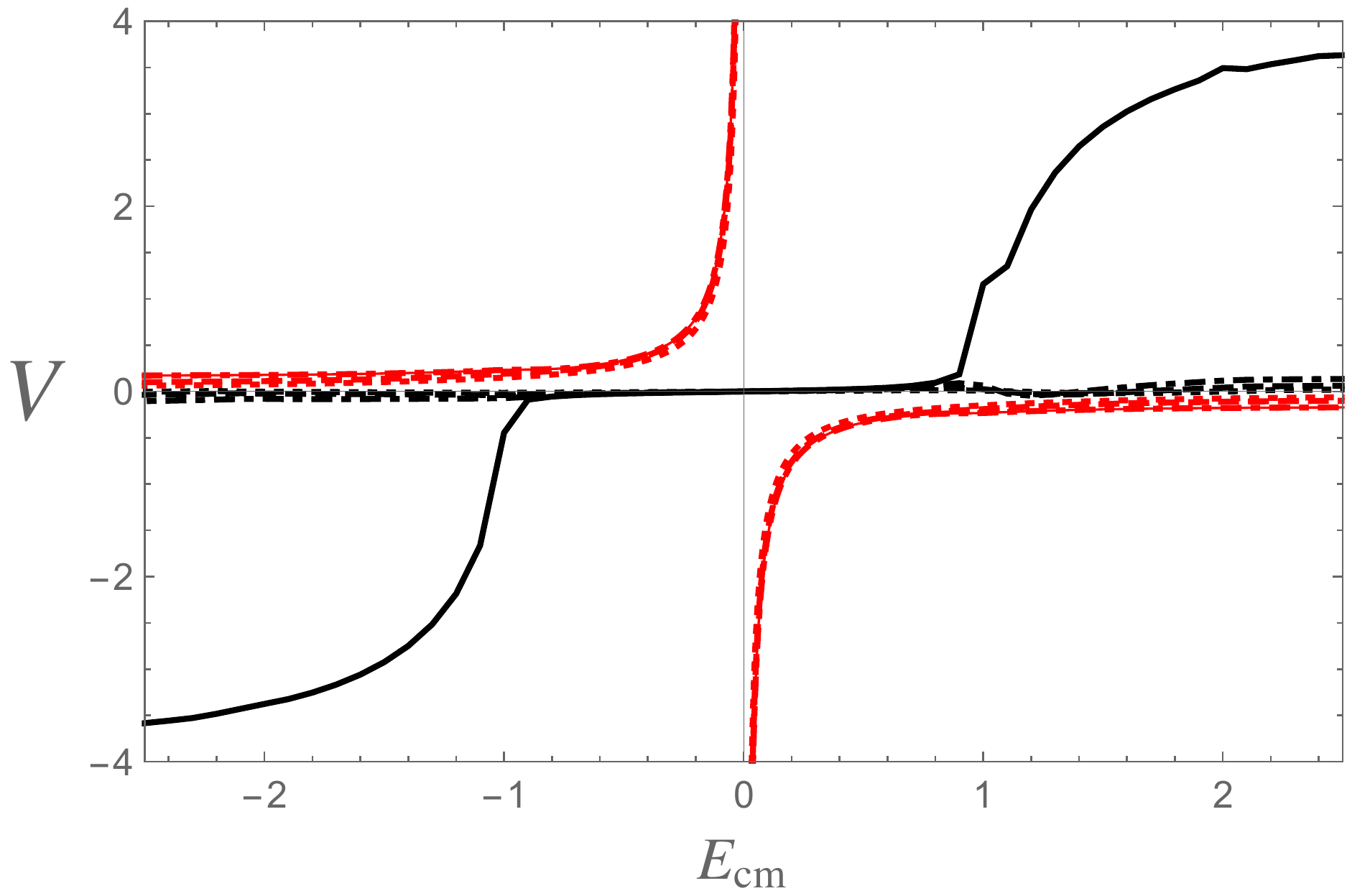}}
  \vspace{-2.3in}
\rightline{\includegraphics[height=2.3in]{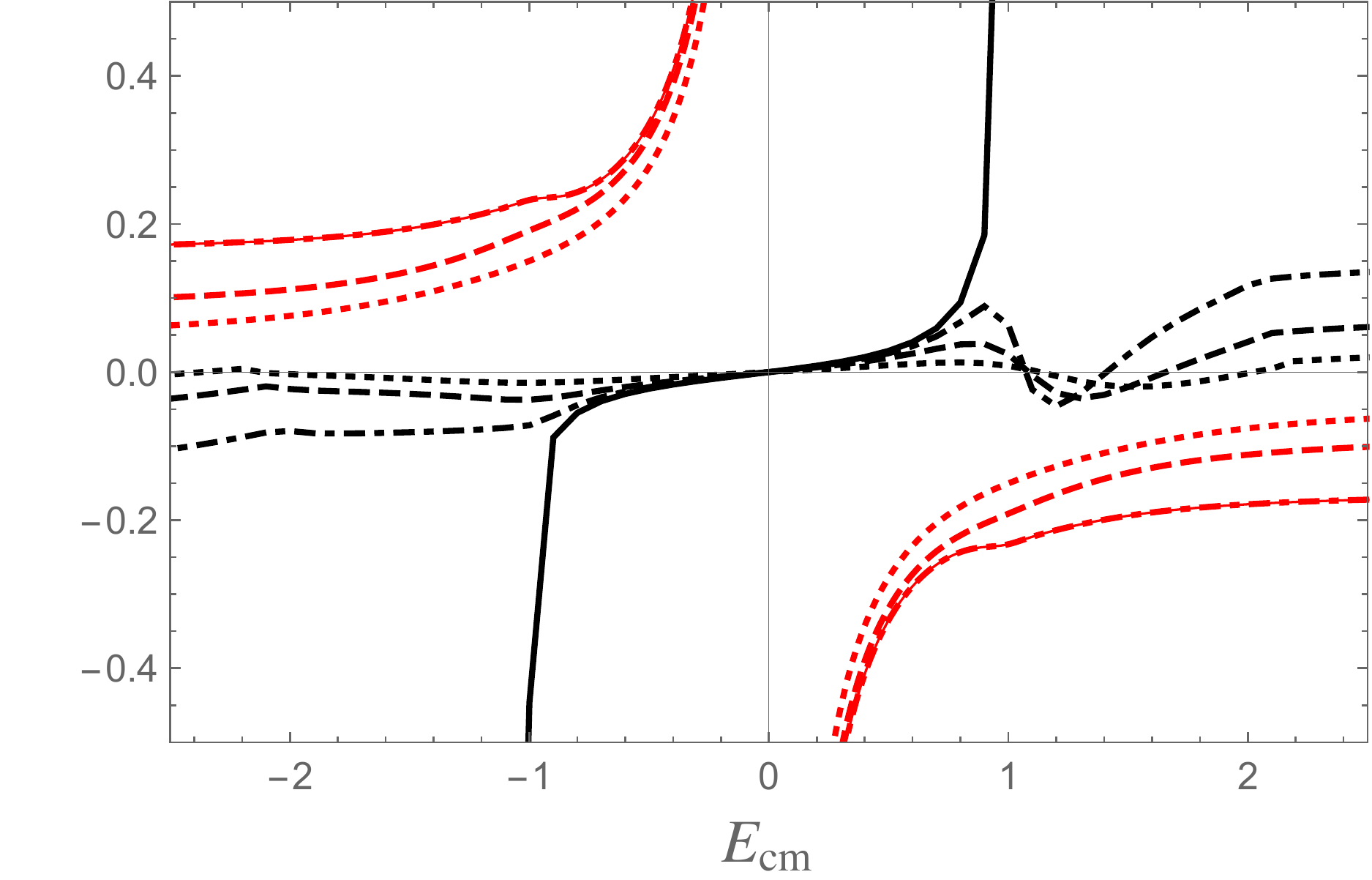}}
 \caption{{\footnotesize The integrated kernels $V=-16V^{4}_{\rm D}/g^4$ (black lines) and $V=-16\,V^{2\times 2}_{\rm D}/g^4$ (red lines) for $\lambda=1$ (dotted), 0.7 (dashed), 0.5 (dot-dashed) and 0.1 (solid) as a function of $E_{\rm cm}$  for the special case when $p=p'=0$ and $\mu=1$.  Left panel:  large vertical scale. Right panel: small vertical scale. Note that, for $-16\,V^{2\times 2}_{\rm D}/g^4$, the cases $\lambda=0.1$ and 0.5 overlap, appearing as a single line.    Units for $E_{\rm cm}$ are the boson mass $\mu$.}}
 \label{fig:V4}
\end{figure*}

Calculation of (\ref{eq:combos2}) directly in the limit ${\bf p}={\bf p}'\to0$ requires evaluation of the double poles at $\kappa=\kappa_\omega^\pm$, giving
\bea
V^{4{\rm D}}_{11}&=&i\int\frac{d^4 k}{(2\pi)^4} \frac{E_{\rm cm}(V^{\rm D}V^{\rm D})_{11}}{8M^2\kappa(\kappa^2-E_{\rm cm}^2)}
=\frac{g^4 M\mu^2}4\,  2\Re\int_k \frac{d}{d\kappa}\Bigg[\frac{E_{\rm cm}(\omega^2-\kappa^2)^2}{\kappa(\kappa^2-E_{\rm cm}^2)(\kappa_\omega^--\kappa)^2\big[I_\omega^2+(R_\omega+\kappa)^2\big]^2}\Bigg]\Bigg|_{\kappa=\kappa_\omega^+}
\nonumber\\
&=&-\frac{g^4 M\mu^2}4\, \Re\int_k\frac{i E_{\rm cm} \big[\omega^2-(\kappa_\omega^+)^2\big] N_{11}}
{4 \lambda_\mu^6 (\kappa_\omega^+)^4\big[(\kappa_\omega^+)^2-E_{\rm cm}^2\big]^2}= -\frac{g^4}{16}\int \frac{d^3k}{(2\pi)^3}\,\widetilde{v}^{4{\rm D}}_{11}
\eea
where
\bea
N_{11}=\big[\omega^2-(\kappa_\omega^+)^2\big]\big[(\kappa_\omega^+)^2-E_{\rm cm}^2\big] (\kappa_\omega^+R_\omega + I_\omega^2 +2 i I_\omega \kappa_\omega^+ ) + \frac12 i\,\lambda_\mu^2\Big(3(\kappa_\omega^+)^2 (\omega^2-E_{\rm cm}^2)+ (\kappa_\omega^+)^4-\omega^2E_{\rm cm}^2\Big). \qquad
\eea
\end{widetext}
Here the $\lambda_\mu^2\to0$ limit is more difficult because both the numerator and denominator must be expanded.  The result is
\bea
\lim_{\lambda_\mu^2\to0} V^{4{\rm D}}_{11}&=&-\frac{g^4 M\mu^2}{8}\int_k \frac{E_{\rm cm}(2\omega^2-E_{\rm cm}^2)}{\omega^4 (\omega^2-E_{\rm cm}^2)^2}\qquad
\eea
in agreement with (\ref{eq:combos2}).  The integral for the total result is
\bea
V^{4}_{\rm D}&\equiv& V^{4{\rm D}}_{11} + V^{4{\rm D}}_{21}
\nonumber\\
&=& -\frac{g^4}{16}\int \frac{d^3k}{(2\pi)^3}\,\big[\widetilde{v}^{4{\rm D}}_{11}+\widetilde{v}^{4{\rm D}}_{21}\big]
\nonumber\\
&=&-\frac{g^4}{16}\int \frac{d^3k}{(2\pi)^3}\,\widetilde{v}^4_{\rm D}\, .
\eea

\begin{figure*}
\leftline{\includegraphics[height=2.3in]{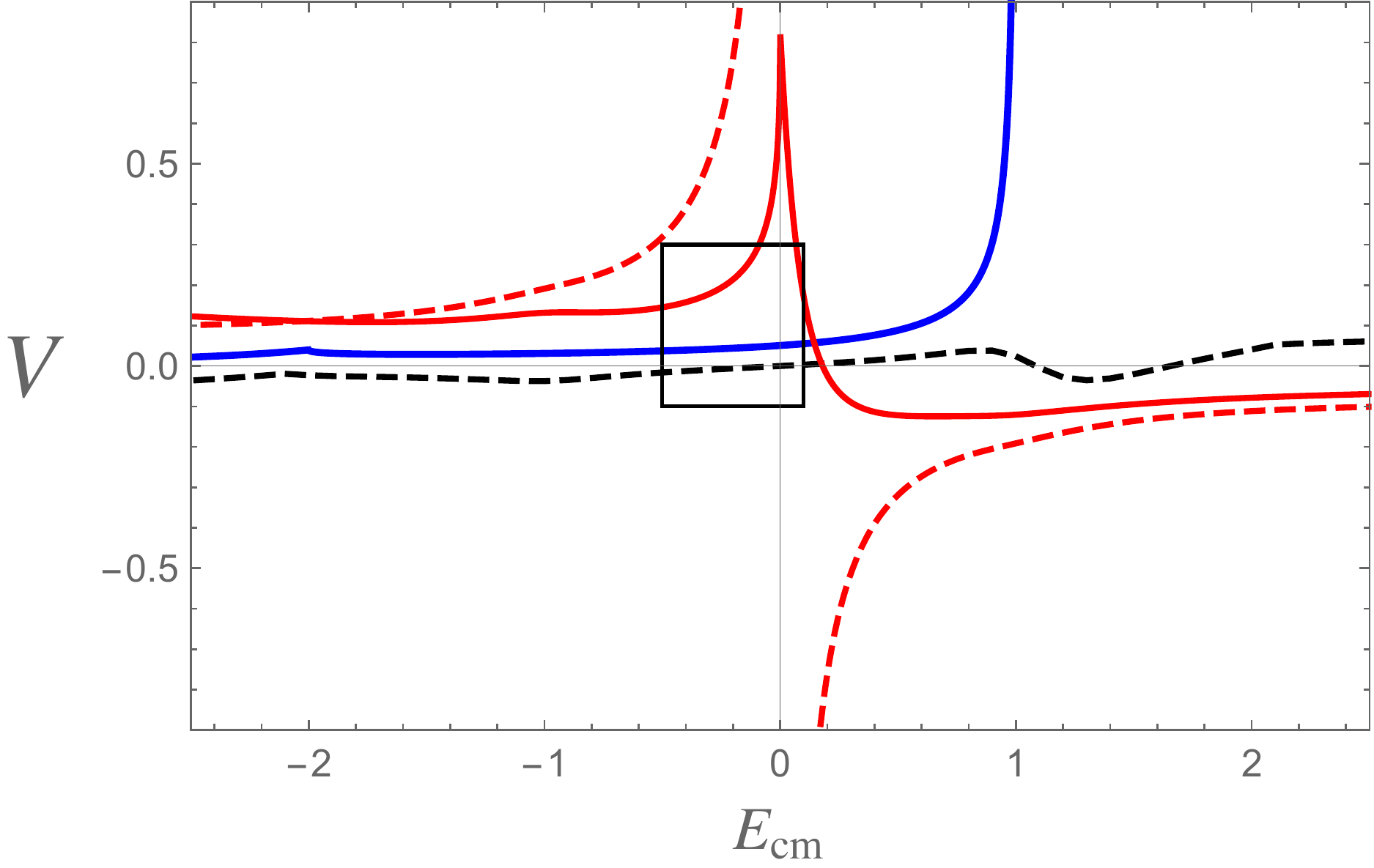}}
  \vspace{-2.3in}
\rightline{\includegraphics[height=2.3in]{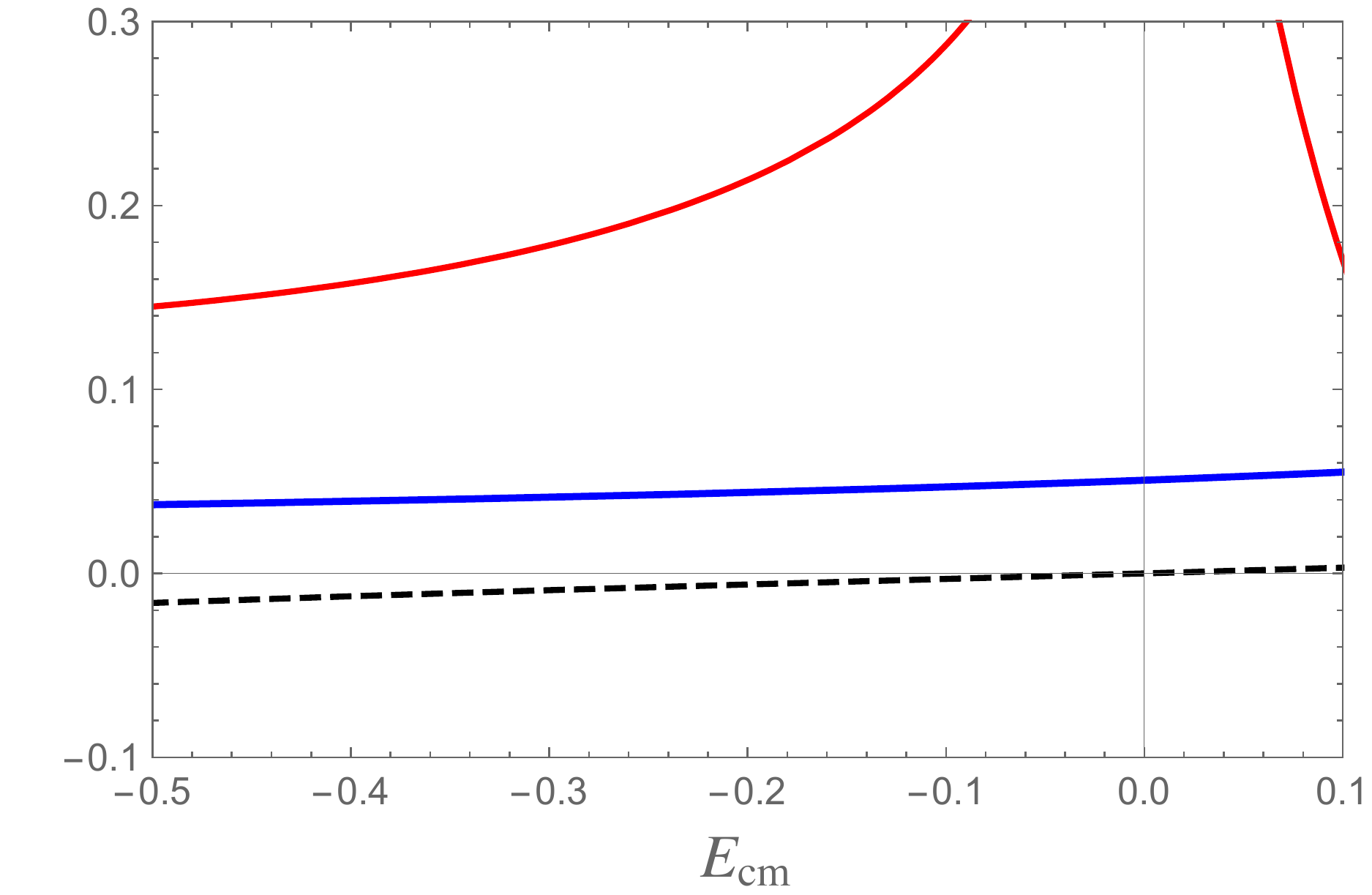}}
 \caption{{\footnotesize Lines showing the integrated kernel $V=-16V^{4}_{\rm BS}/g^4$  (blue), and for $\lambda_\mu=0.7$, the kernels  $V=-16 V^{4}_{\rm D}/g^4$  (black dashed), $V=-16\,(\Re V^{2\times 2}_{\rm D})/g^4$ (red), $V=-16\,V^{2\times 2}_{\rm D0}/g^4$ (red, dashed). Left panel shows these on a large scale; right panel gives the details inside the box shown in the left panel.   Units for $E_{\rm cm}$ are the boson mass $\mu$.}}
 \label{fig:V407}
\end{figure*}

\begin{figure}
\leftline{\includegraphics[width=3.3in]{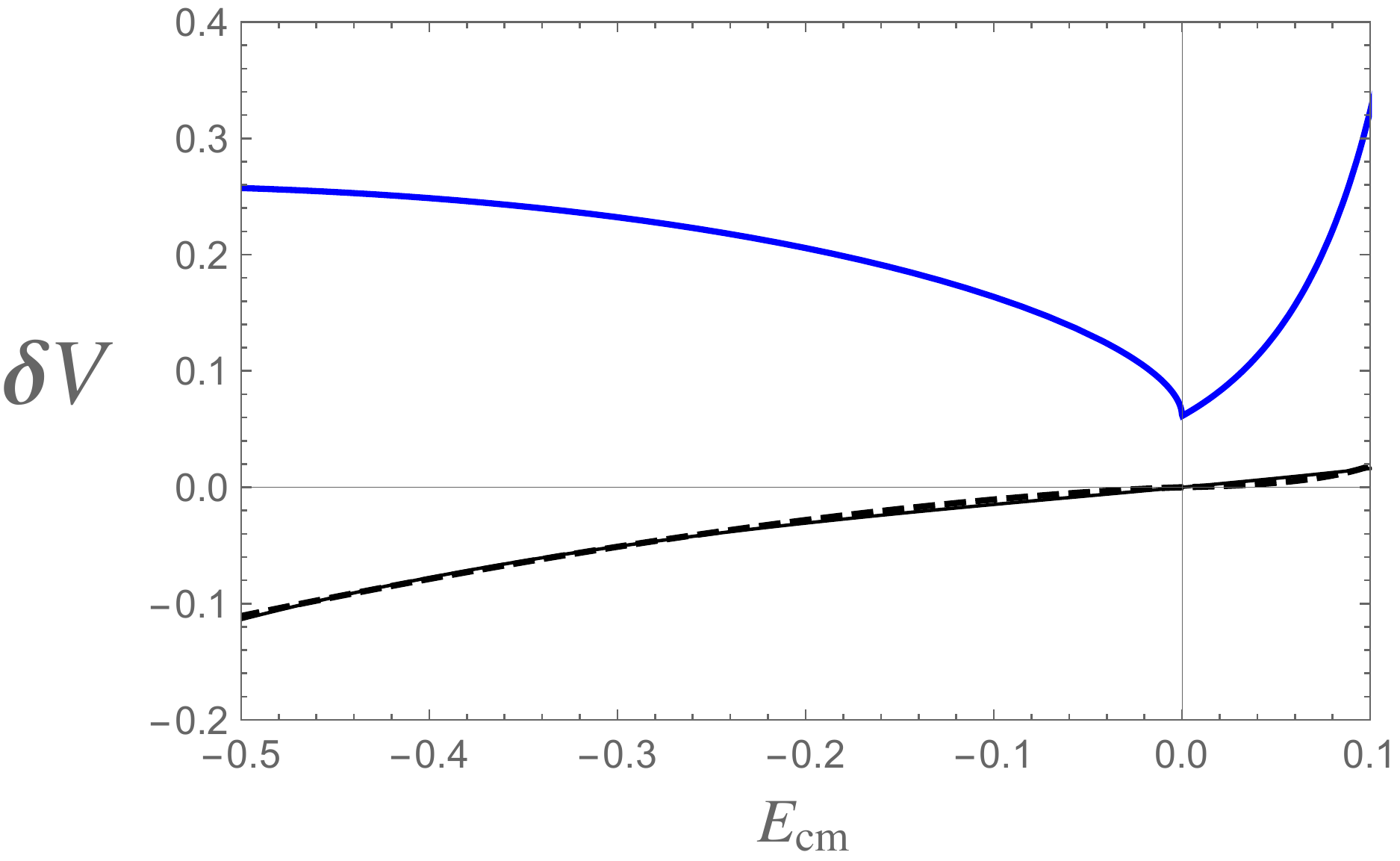}}
 \caption{{\footnotesize Lines showing, for $\lambda_\mu=0.7$ the ratio $\delta V^4_{\rm BS}=V^{4}_{\rm BS}/V^{2\times 2}_{\rm D}$ (blue), and $\delta V^4_{\rm D}=V^{4}_{\rm D}/V^{2\times 2}_{\rm D}$ (black dashed). The thin black line which fits $\delta V^4_{\rm D}$ is discussed in the text.  Units for $E_{\rm cm}$ are the boson mass $\mu$.}}
 \label{fig:deltaV4}
\end{figure}

Fig.~\ref{fig:tildeV} shows the integrands $\widetilde v$ as a function of $E_{\rm cm}$ for values of $\lambda=0.1,\, 0.5, \,0.7$, and 1, for the special case when $\mu=1$ and all three-momenta are zero.   The same figure also shows $\widetilde{v}^{4}_{\rm BS}$, defined in Eq.~(\ref{eq:BS}), and $\widetilde{v}^{4}_{\rm B}$, defined Eq~(\ref{eq:cancellation2}). 

The overlap of the lines for $\widetilde{v}^{4}_{\rm B}$ and 
$\widetilde{v}^{4}_{\rm D}$ for $\lambda=0.1$ show that the results for small $\lambda_\mu$ converge smoothly to Model B.  The line for  $\widetilde{v}^{4}_{\rm BS}$ shows no instability singularity, but the production singularity remains (because no form factor is used in the BS calculation).  Note that, in the region of moderate $-0.5\mu\leq E_{\rm cm}\leq 0.5\mu$, the integrand for the fourth order BS kernel is much larger than the CST kernels, which vanish at $E_{\rm cm}=0$ because of the cancellation theorem.   As anticipated, both the CST instability and production singularities are removed when the form factor is added, and as $\lambda$ increases the integrand shows less and less structure in these regions.   

Finally, note that the 2-boson singularities at  $E_{\rm cm}=\pm2 R_\omega \simeq \pm2\mu$ are present in {\it all\/} cases.  As discussed above, these are a feature of the fourth-order integrands that, through the principal value prescription, will still give finite fourth order kernels.

\subsection{Comparison of the first iteration of OBE with  irreducible fourth order kernels}

I conclude this section with a discussion of the energy dependence of the irreducible CST fourth order kernel, $V^4_{\rm D}$, the first iteration of the the OBE,  both the {\it real\/} part of the exact result $V^{2\times2}_{\rm D}$ (the elastic cut starts at $E_{\rm cm}=0$) given in Eq.~(\ref{eq:2by2}) and the approximate result $V^{2\times2}_{\rm D0}$ (real everywhere) given in Eq.~(\ref{eq:2by2app}), and the irreducible BS kernel $V^4_{\rm BS}$ defined in Eq.~(\ref{eq:BS}), all evaluated at external momenta $p=p'=0$.  To obtain these results the integrands for each are integrated over the internal three momentum $k$. The integrand for $V^{2\times2}_{\rm D0}$ is a smooth function easily evaluated, and the real part of $V^{2\times2}_{\rm D}$ is evaluated using the principal value prescription to treat the elastic scattering singularity from the the propagator.   The integrands for $V^4_{\rm D}$ were shown in Fig.~\ref{fig:tildeV} for $k=0$, and, because of the form factor, have singularities only at $E_{\rm cm}=\pm2 R_\omega$ which can be evaluated using  the principle value prescription.  The integral for $V^4_{\rm BS}$ is defined only for $E_{\rm cm}<\mu$.  Figs.~\ref{fig:V4} and \ref{fig:V407} show the results. 

Fig.~\ref{fig:V4} shows that the choice $\lambda_\mu=0.1$ gives a very large kernel for $|E_{\rm cm}|>1$, so full control the instability and production singularities in $V^4_{\rm D}$ requires a choice $\lambda_\mu\gtrsim0.5$. With these choices  the iteration of the OBE remains significantly larger than the irreducible fourth order kernel, justifying the conclusion that, even though the cancellation theorem is violated except at very small energy,  the Model D treatment of OBE still accurately approximates the description of the physics of the generalized ladder sum to fourth order.  

Fig.~\ref{fig:V407}, focusing on the moderate choice $\lambda_\mu=0.7$,  shows the comparative sizes of $V^{2\times2}_{\rm D}, V^4_{\rm D}$, and $V^4_{\rm BS}$.  Note that the exact real part of $V^{2\times2}_{\rm D}$ has a zero around $E_{\rm cm}\sim 0.2\mu$; in this region the imaginary part is dominate and realistic conclusions cannot be drawn without including it.  (Because of this, I focus the discussion on the energy region $-0.5\mu\leq E_{\rm cm}\leq 0.1\mu$.)  The fourth order CST kernel is much smaller than the BS kernel.   This is emphasized  in Fig.~\ref{fig:deltaV4} showing the ratio of  the BS and Model D kernels to the real part of exact $V^{2\times2}_{\rm D}$ (for $\lambda_\mu=0.7$).

I conclude that the OBE approximation to the CST is more accurate than the ladder approximation to the BS.  This result has already been proved for non-identical particles, but this is the first time it has been proved for equal masse particles when symmetries are important.

 \section{Conclusions} \label{sec:conclusions} 
 
 What I now refer to as the CST was introduced over 50 years ago and has been widely used to describe $q \bar q$ bound states, $NN$ scattering, deuteron and pion form factors, and many other systems.  One of its shortcomings was the presence of unphysical instability singularities for total energies $W\leq 2m-\mu$ that automatically arise when the one-boson-exchange CST equations are symmetrized for the treatment of systems of identical particles, or the treatment of $q \bar q$ bound states with charge conjugation symmetry.  I emphasize that the principal issue with these singularities is a theoretical one: while they lead to finite results in numerical calculations they are an unpleasant sign that something is missing and leave doubts that the major physics is under control.
 
 This paper shows how these singularities can be removed by introducing a form factor of the type given in Eq.~(\ref{eq:Fq2}), which depends only on $q^2$ and one parameter $\lambda$.  Any choice of $\lambda>0$ will eliminate the singularity, but the results presented here suggest that the choice $\lambda\simeq 0.7$ is large enough to smooth out the rapid behavior of the kernel in the region of the former singularity and also small enough to preserve the behavior of the original OBE kernel away from this region.  In particular, for on-shell scattering (where $q^2\leq 0$ always holds), the modifications introduced by the form factor are small.  Other form factors can be added to improve convergence, if needed.   While all specific calculations in this paper were limited to spinless theories of the $\phi\psi^*\psi$ type, the use of a form factor to eliminate singularities can easily be extended to theories with particles that have non-zero spins and isospins.

Conclusions about the accuracy of the OBE approximation, however, depend more specifically on the details of the theory. For theories of the $\phi\psi^*\psi$ type discussed in this paper, Fig.~\ref{fig:deltaV4} summarizes the errors that result from the omission of the irreducible $V^4$ kernels.  In the BS theory, they lead to errors as large as 10\% to 30\%, while in the CST they are much smaller, varying from 0 to -10\% over a range of center of mass energies varying from 0 to $-0.5\mu$.  The fit shown in Fig.~\ref{fig:deltaV4} gives 
\bea
\delta V^4_{\rm D}\simeq 0.15 \frac{E_{\rm cm}}{\mu}+0.07\left(\frac{E_{\rm cm}}{\mu}\right)^2+0.45\left(\frac{E_{\rm cm}}{\mu}\right)^3 . \qquad
\, .  \label{eq:deltaV}
\eea   
Even though my estimates ignore the dependence of $\delta V^4$ on the external momenta $p$ and $p'$, the center of mass energy, if due to the momentum of the nucleons or quarks could be taken to be of order $p^2/M$, so that (\ref{eq:deltaV}) suggests that the error in the CST ladder sum is of order $p^2/M^2$ and higher, very small indeed.

In conclusion: when the CST is applied to a system of two identical spin zero ``nucleons'' (or a flavor neutral spin zero ``quark-antiquark'' pair)  exchanging spin zero bosons, the cancellation theorem (originally proved for systems of non-identical particles) is violated by the exchange terms required to ensure the necessary symmetry.  However, removing the singularities carried by these exchange terms reduces the violation of the cancellation theorem to a very small effect.  Just as in the case of non-identical particles, a better ladder approximation (i.e. OBE approximation) to the sum of generalized ladders is obtained by using the CST, rather than the  BS equation.  

Do these cancellations work for more realistic theories?   In particular, can one justify a OBE approximation for the treatment of the $NN$ interaction? Modern studies using the algebra of large $N_c$ QCD are promising; see the work  of Ref.~\cite{Banerjee:2001js}.

With this insight, one of the main issues with the CST has been eliminated and the way is clear for further applications.



\acknowledgements

This work was partially supported by Jefferson Science
Associates, LLC, under U.S. DOE Contract No.~DE-AC05-
06OR23177.   It is a pleasure to thank J.~W.~Van Orden, Alfred Stadler, and Carl Carlson for their significant contributions to the development of the CST over the years.   I also thank the Portuguese group (Elmar Biernat, M. T. Pe\~na, and Alfred Stadler) for recent and ongoing work on quark self-energies that stimulated work this paper.

\vspace*{0.2in}

\appendix

\begin{widetext}

\section{Proof that the CST subtracted box has no scattering singularities} \label{app:A}

Using the OBE defined in (\ref{eq:OBE}),  the iterated OBE, consisting of  both direct and exchange terms, becomes
\bea
V^{2\times 2}_{\rm dir}(\hat p,\hat p')= -\frac12&&\int_k V(\hat p,\hat k) G(\hat k)
\Big\{V(\hat k,\hat p')+ V(P-\hat k,\hat p')\Big\}
\nonumber\\
V^{2\times 2}_{\rm ex}(\hat p,\hat p')=-\frac12&&\int_k V(P-\hat p,\hat k) G(\hat k)
\Big\{V(\hat k,\hat p')+ V(P-\hat k,\hat p')\Big\} 
\label{eq:OBEiterated}
\eea
where this is the first iteration of  Eq.~(\ref{eq:2channel}).  
Subtracting this from 
the box contributions  [(\ref{eq:Vdir0}) and (\ref{eq:Vex0}) with $\eta=0$], 
gives
  \bea
{V}_{\rm dir}^{4{\rm box}}&=& {M}_{\rm dir}^{4{\rm box}}-V^{2\times 2}_{\rm dir}
\nonumber\\
&=&\frac{i}{2}\int\frac{d^4k}{(2\pi)^4}\Bigg[\frac1{M^2-k^2-i\epsilon} -\frac{\pi i\,\delta(E_k-k_0)}{E_k}\Bigg]\frac{V(\hat p,k)\,\Big\{V(k,\hat p')+V(P-k,\hat p')\Big\}}{M^2-(P-k)^2-i\epsilon} 
\nonumber\\
{V}_{\rm ex}^{4{\rm box}}&=& {M}_{\rm ex}^{4{\rm box}}-V^{2\times 2}_{\rm ex}
\nonumber\\
&=&\frac{i}{2}\int\frac{d^4k}{(2\pi)^4}\Bigg[\frac1{M^2-k^2-i\epsilon} -\frac{\pi i\,\delta(E_k-k_0)}{E_k}\Bigg] \frac{V(P-\hat p,k)\,\Big\{V(k,\hat p')+V(P-k,\hat p')\Big\}}{M^2-(P-k)^2-i\epsilon} ,
  \qquad\;
\eea
showing explicitly how the $V^{2\times2}$ terms cancel; the  pole at $k_0=E_k$, completing the proof.  

For future reference, the explicit form of the iterated OBE contributions in the limit ${\bf p}, {\bf p}'\to 0$, as as discussed in Sec.~\ref{sec:m4andv4}, is
\bea
V^{2\times2}_{{\rm D}0} = \frac{g^4}{16}\int\frac{d^3k}{(2\pi)^3}\widetilde{v}^{4\,{\rm D}}_{2\times 2}=\frac{g^4\mu^2}{16 E_{\rm cm}}\int\frac{d^3k}{(2\pi)^3}\Bigg[\frac{\omega^2}{(\lambda_\mu^4 +\omega^4)}+\frac{\omega^2-E_{\rm cm}^2}{\lambda_\mu^4 +(\omega^2-E_{\rm cm}^2)^2}\Bigg]^2\, . \label{eq:2by2app}
\eea
Note that when $\lambda_\mu\to0$, this reduces to the limiting results of the sum of the $\kappa=0$ pole contributions in Eqs.~(\ref{eq:M11box})--(\ref{eq:M22box}).  However, it is easy to calculate this leading term exactly, which modifies (\ref{eq:2by2app}), particularly for small  $E_{\rm cm}$.  Dropping the $i\epsilon$ prescription, which produces the elastic cut, gives
\bea
V^{2\times2}_{{\rm D}} = -\frac{g^4}{16}\int\frac{d^3k \;M}{(2\pi)^3 E_k}\left[\frac{2\mu^2}{W(2 e_k-E_{\rm cm})}\right]\Bigg\{\frac{\omega^2- e_k^2}{\lambda_\mu^4 +(\omega^2- e_k^2)^2}+\frac{\omega^2-(E_{\rm cm}-e_k)^2}{\lambda_\mu^4 +(\omega^2-(E_{\rm cm}-e_k)^2)^2}\Bigg\}^2\, , \label{eq:2by2}
\eea
where $e_k=E_k-M$.  Note that $e_k\to0$ and (\ref{eq:2by2}) converges to (\ref{eq:2by2app}) as $M\to\infty$.


\section{Illustrative integrals}\label{app:B}


Consider the integral $I_1$ defined below, where $dk_0\to d\kappa$ and $\omega^2=\mu^2+k^2$.  Closing the contour in the lower half complex plane:  
\bea
 I_1&=&i\int\frac{d^3k}{(2\pi)^3}\int_{-\infty}^\infty \frac{d\kappa}{2\pi}\frac1{\underbrace{(\omega+\kappa-i\epsilon)}_{1}\underbrace{(\omega-\kappa-i\epsilon)}_{2}
 \underbrace{(\omega-E_c+\kappa-i\epsilon)}_{3}\underbrace{(\omega+E_c-\kappa-i\epsilon)}_{4}}
 \nonumber\\
 &=&-\int\frac{d^3k}{(2\pi)^3}  
 \Bigg[\frac1{2\omega E_c(2\omega-E_c-i\epsilon)} 
 - \frac1{2\omega E_c(2\omega+E_c-i\epsilon)}\Bigg].\qquad
 \eea 
 \end{widetext}
This displays the fact that the production singularity comes from a pinch between pole 2 in the lower half plane and  pole 3  in the upper half plane, while the instability singularity comes from a pinch between pole 4 in the lower half plane and  pole 1 in the upper half plane.   Both of these  singularities are real and a property of the integral.  

An alternative discussion of the integral shows how the singularities give rise to cuts.  Use the Feynman parameterization, and using $d^4k=d^3kd\kappa$, $I_1$ becomes
\bea
I_1&=&
i\int \frac{d^4k}{(2\pi)^4}\int_0^1 dz\frac{1}{\big[\omega^2-\kappa^2+z(2\kappa E_c-E_c^2)-i\epsilon\big]^2}
\nonumber\\
&=&i\int \frac{d^3k}{(2\pi)^4}\int_0^1dz \int_{-\infty}^\infty\frac{d\kappa'}{\big[\omega^2-\kappa'^2-z(1-z)E_c^2-i\epsilon\big]^2}
\nonumber\\ &&
\eea
where $\kappa'=\kappa+zE_c$.  Now, assuming that 
$4\mu^2-E_c^2>0$, and introducing $\eta=4\omega^2$, two of the integrals are evaluated,  allowing me to cast the result into a dispersive form
\bea
I_1&=&\int \frac{d^3k}{4(2\pi)^3}\int_0^1dz \frac{1}{\big[\omega^2-z(1-z)E_c^2-i\epsilon\big]^{3/2}}
\nonumber\\
&=&\int \frac{d^3k}{(2\pi)^3}\frac{1}{\omega\big[4\omega^2-E_c^2-i\epsilon\big]}
\nonumber\\
&=&\frac1{4\pi^2}\int_{4\mu^2}^\infty \frac{d\eta \rho(\eta)}{\eta-E_c^2-i\epsilon}\, ,
\eea
where the familiar two-body phase space factor is 
\bea
\rho(\eta)=\sqrt{\frac{\eta-4\mu^2}{\eta}}\, .
\eea
While this integral was initially evaluated under the assumption that $E_c^2\leq4\mu^2$, the final result is a dispersion integral showing that it can clearly be extended to larger values of $E_c^2$, where a two-boson production singularity appears for $E_c\geq 2\mu$ and an instability 
for $E_c     \leq 2\mu$.

\section{2-boson singularities in $V^4_{\rm D}$} \label{app:sing}

The 2-boson singularities arise from the contribution $V^{4{\rm D}}_{21}$  defined in Eq.~(\ref{eq:V21D}).  Multiplying the numerator and denominator by the complex conjugate of the denominator, the absolute value squared of each term in the denominator can be examined for zeros.  The only term that leads to such a zero is the term involving $\lambda_\mu^4$.  To simplify the discussion I take $p, p' \to0$, and denote $\omega=\sqrt{\mu^2+k^2}$.  Then these terms become 
\bea
|d_4^\pm|^2&=&\big\{\lambda_\mu^4+[\omega^2-(R_\omega+i I_\omega\pm E_{\rm cm})^2]^2\big\}
\nonumber\\&&
\times\big\{\lambda_\mu^4+[\omega^2-(R_\omega-i I_\omega\pm E_{\rm cm})^2]^2\big\}
\nonumber\\
&=&{\cal R}_4^2+{\cal I}_4^2
\eea
where
\bea
{\cal R}_4 &=& \lambda_\mu^4+\big[\omega^2-(R_\omega\pm E_{\rm cm})^2+I_\omega^2\big]^2-4I_\omega^2(R_\omega\pm E_{\rm cm})^2
\nonumber\\
{\cal I}_4&=&4(R_\omega\pm E_{\rm cm})\big[\omega^2-(R_\omega\pm E_{\rm cm})^2+ I_\omega^2\big]\, .
\eea
Since $|d_4^\pm|^2$ is the sum of two squares, one might think it cannot be zero, but it is possible for both terms to be zero at the same time.  Using (\ref{eq:subsdefs}) to replace $I_\omega$, and noting that $\omega^2 R_\omega^2+\lambda_\mu^4/4=R_\omega^4$, this occurs when
\bea
0&=&\omega^2-(R_\omega\pm E_{\rm cm})^2+ I_\omega^2
\nonumber\\
&=&\omega^2-(R_\omega\pm E_{\rm cm})^2+ \frac{\lambda_\mu^4}{4R_\omega^2}\to
R_\omega^2-(R_\omega\pm E_{\rm cm})^2
\nonumber\\
\nonumber\\
0&=&\lambda_\mu^4-4I_\omega^2(R_\omega\pm E_{\rm cm})^2\to 1-\frac{(R_\omega\pm E_{\rm cm})^2}{R_\omega^2}\, .
\eea 
Hence both equations are solved when
\bea
E_{\rm cm}&=&\mp\, 2R_\omega \quad{\rm or}\quad 0\, .
\eea
These are a generalization of the 2-boson singularities discussed in Sec.~\ref{sec:m4andv4}.

\end{document}